\documentclass[%
 reprint,
nofootinbib,
 amsmath,amssymb,
 aps,prd,
floatfix,
]{revtex4-2}

\usepackage{graphicx}
\usepackage{dcolumn}
\usepackage{bm}
\usepackage{hyperref}
\usepackage{color}
\usepackage{comment}
\usepackage{slashed}
\usepackage{subfigure}
\usepackage{multirow}
\usepackage{braket}
\usepackage[normalem]{ulem}
\usepackage{tikz}
\usepackage[compat=1.1.0]{tikz-feynman}
\usepackage{ragged2e}
\usepackage{tikz} 
\usetikzlibrary{shapes,arrows,positioning,automata,backgrounds,calc,er,patterns, arrows, circuits.ee.IEC, positioning, calc}
\usepackage[compat=1.1.0]{tikz-feynman}
\usepackage[american, oldvoltagedirection, american currents,siunitx]{circuitikz}
\usepackage{booktabs}
\usepackage{url} 
\usepackage[version=3]{mhchem}
\usepackage{fontawesome} 
\usepackage{xspace}
\usepackage{braket}

\definecolor{linkcolor}{rgb}{0.7752941176470588, 0.22078431372549023, 0.2262745098039215}

\newcommand{\nbicon}{{\color{linkcolor}\faFileCodeO}\xspace}
\newcommand{\nblink}[1]{\href{https://github.com/kenvantilburg/piezoaxionic-effect/blob/main/code/#1}{\nbicon}}
\newcommand{\githubmaster}{\href{https://github.com/kenvantilburg/piezoaxionic-effect/}{\faGithub}\xspace}

\definecolor{deepgreen}{rgb}{0.2,0.8,0.2}

\definecolor{deepblue}{rgb}{0.2,0.2,0.8}

\definecolor{deepred}{rgb}{0.8,0.2,0.2}

\newcommand{\vect}[1]{\boldsymbol{\mathbf{#1}}}
\newcommand{\half}{\frac{1}{2}}
\newcommand{\threehalf}{\frac{3}{2}}

\newcommand{\dd}{{\rm d}}
\newcommand{\cc}{\text{c.c.}}
\newcommand{\Sch}{\vect{\mathsf{S}}}

\makeatletter
\def\l@subsubsection#1#2{} 
\makeatother

\begin{document}


\title{The Piezoaxionic Effect}
\author{Asimina Arvanitaki}
\email{aarvanitaki@perimeterinstitute.ca}
\affiliation{Perimeter Institute for Theoretical Physics, Waterloo, Ontario N2L 2Y5, Canada}

\author{Amalia Madden}
\email{amadden@perimeterinstitute.ca}
\affiliation{Perimeter Institute for Theoretical Physics, Waterloo, Ontario N2L 2Y5, Canada}
\affiliation{Department of Physics and Astronomy, University of Waterloo, Waterloo, Ontario, N2L 3G1, Canada}

\author{Ken Van Tilburg}
 \email{kenvt@nyu.edu}
 \email{kvantilburg@flatironinstitute.org}
 \affiliation{Center for Cosmology and Particle Physics, Department of Physics, New York University,
New York, NY 10003, USA}
 \affiliation{Center for Computational Astrophysics, Flatiron Institute, New York, NY 10010, USA}
 
\date{\today}

\begin{abstract}
Axion dark matter (DM) constitutes an oscillating background that violates parity and time-reversal symmtries. Inside piezoelectric crystals, where parity is broken spontaneously, this axion background can result in a stress. We call this new phenomenon ``the piezoaxionic effect". When the frequency of axion DM matches the natural frequency of a bulk acoustic normal mode of the piezoelectric crystal, the piezoaxionic effect is resonantly enhanced and can be read out electrically via the piezoelectric effect. We explore all axion couplings that can give rise to the piezoaxionic effect---the most promising one is the defining coupling of the QCD axion, through the anomaly of the strong sector. We also point our another, subdominant phenomenon present in all dielectrics, namely the ``electroaxionic effect". An axion background can produce an electric displacement field in a crystal which in turn will give rise to a voltage across the crystal. The electroaxionic effect is again largest for the axion coupling to gluons. We find that this model independent coupling of the QCD axion may be probed through the combination of the piezoaxionic and electroaxionic effects in piezoelectric crystals with aligned nuclear spins, with near-future experimental setups applicable for axion masses between $10^{-11}\,\mathrm{eV}$ and $10^{-7}\,\mathrm{eV}$, a challenging range for most other detection concepts.
\end{abstract}

\maketitle

\tableofcontents

\section{Introduction}\label{sec:intro}

Axions are among the best motivated extensions of the Standard Model. In particular, the quantum chromodynamics (QCD) axion is a natural consequence of the symmetry that can explain the smallness of the neutron's electric dipole moment and the absence of other charge-parity-violating phenomena in the strong interactions~\cite{Wilczek:1977pj,Weinberg:1977ma,Peccei:1977hh,Dine:1982ah}, providing a solution to the strong $\mathsf{CP}$ problem. In string theory, a plethora of axions is a natural consequence of topological complexity in compactified dimensions~\cite{Svrcek:2006yi,Arvanitaki:2009fg}. At the same time, axions are excellent dark matter (DM) candidates, since the misalignment mechanism of production ensures a consistent cosmological history, an indispensable ingredient of any DM model~\cite{Preskill:1982cy,abbott1983cosmological,Dine:1982ah}. 

When axions make up (all or a component of) the DM of our Universe, they manifest as a background that violates parity ($\mathsf{P}$) and time reversal ($\mathsf{T}$) invariance.
In this paper, we show how a $\mathsf{P}$-violating axion DM background produces a stress in piezoelectric crystals, a new observable that we call the \emph{piezoaxionic effect}. Piezoelectric crystal structures break parity, by definition. Therefore, no symmetry forbids the occurrence of stress (even under $\mathsf{P}$ and $\mathsf{T}$) upon application of an electric field (odd under $\mathsf{P}$ and even under $\mathsf{T}$)---the converse piezoelectric effect~\cite{Lippmann,CurieJP3}. Vice versa, an electric field results from an applied stress---the piezoelectric effect~\cite{CurieJP1,CurieJP2}.
For such a material, a stress can appear across the crystal in an axion DM background, by analogy to the (converse) piezoelectric effect. For the axion coupling to anomaly of the strong interactions, which violates simultaneously $\mathsf{P}$ and $\mathsf{T}$, a piezoelectric crystal with polarized nuclear spins is needed for the stress to appear. When the oscillation frequency of the axion DM wave matches an acoustic resonance frequency of the crystal, the axion-induced stress results in a resonantly enhanced strain.

Such strain, or simply put, a change in the length of a crystal, is an observable that has been used for decades in the context of resonant-mass gravitational wave detectors~\cite{Weber:1960zz,Weber:1967jye,Aguiar:2010kn}. Resonant-mass GW detectors measure \emph{absolute} changes in length much more more precisely than LIGO~\cite{IGEC-2:2007alz}. A subset of the authors in this paper has shown in the past that such sensitivity can be used to probe scalar DM~\cite{arvanitaki2016sound}. The AURIGA collaboration implemented this idea and used their existing data to place the most stringent constraints thus far on ($\mathsf{P}$- and $\mathsf{T}$-even) scalar DM that couples to the electron and/or the photon~\cite{Branca:2016rez}, showing sensitivity to strains as small as $10^{-25}$, provided the signal has a sufficiently long coherence time.

We expand on these ideas and demonstrate how a resonant-mass detector made from a piezoelectric crystal can be used to detect QCD axion DM with masses between $10^{-11}\,\mathrm{eV}$ and $10^{-7}\,\mathrm{eV}$. While several axion couplings may produce a signal in our setup, the most promising one at low frequencies is the irreducible coupling of the QCD axion to gluons.

There is one more observable associated with the $\mathsf{P}$- and $\mathsf{T}$-violating nature of an axion background. Aligned nuclear spins in any dielectric material, bathed in QCD axion DM, source electric displacements~\cite{Graham:2013gfa}, that produce an oscillating voltage difference across two opposing surfaces of the material. We call this phenomenon the \emph{electroaxionic effect}. As we will see, this effect does not benefit from the same enhancement through mechanical resonance as the piezoaxionic effect, but nevertheless contributes to the axion signal when using an electrical readout. A similar electroaxionic effect is also present for derivative couplings of axion-like particles.

The observables discussed in this paper differ significantly from solid-state EDM experiments~\cite{lamoreaux2002solid,mukhamedjanov2005suggested,budker2006sensitivity, sushkov2010prospects,rushchanskii2010multiferroic} as well as nuclear-magnetic-resonance-based techniques for QCD axion DM detection~\cite{budker2014proposal}, since they do not require the application of an electric field to be effective or the use of ferroelectric materials. Furthermore, both of these types of experiments rely on detecting a change in the magnetization of the sample due to a reversal of the electric field (solid-state EDM experiments) or due to the precession of the spins in the axion background. The piezoaxionic and electroaxionic effects produce an oscillatory signal without any change in the direction of the aligned spins.

Our paper is structured as follows. In Sec.~\ref{sec:theory}, we calculate the piezoaxionic and the electroaxionic tensors in piezoelectric crystals, focusing on the axion coupling to the QCD anomaly. In Sec.~\ref{sec:experiment}, we describe the experimental setup, delineate the signal, discuss noise sources and possible backgrounds, and present our forecast for the reach to axion parameter space. Finally, we present estimates for other couplings in Sec.~\ref{sec:othercouplings} and discuss our assumptions and future directions in Sec.~\ref{sec:discussion}. 
The data and code used to obtain the results of this study is available on GitHub \githubmaster, and a link (\nbicon) below each figure provides the code with which it was generated. 
We use natural units wherein $\hbar = c = k_B = 1$.
For the convenience of the reader, Tab.~\ref{key} summarizes the main symbols introduced in Secs.~\ref{sec:theory} and~\ref{sec:experiment}.
\begin{table}[ht]
    \begin{tabular}{l l c c c}
        \hline \hline
        Quantity         						& Symbol		   & $\mathsf{P}$ & $\mathsf{T}$   & Subsection\\
        \hline
        Axion mass, decay constant              & $m_a$, $f_a$                          &       &       & \multirow{4}{*}{\ref{sec:axion}} \\
        Axion frequency                         & $f$                                   &       &       & \\
        Axion coherence time                    & $\tau_\mathrm{coh}$                   &       &       & \\
        Effective theta angle (with axion)      & $\overline{\theta}_a$                 & $-$   & $-$   & \\
        \hline
        Schiff moment (magnitude)				& $\mathsf{S}$							&  $-$     &   $-$    & \multirow{3}{*}{\ref{sec:nucleus}}\\
        Schiff moment									&$\vect{\mathsf{S}}$ & $-$ & $+$ \\
        Nuclear spin                            & $I_\alpha$                            & $+$   & $-$   & \\         
        \hline
        Atomic charge, number                   & $Z$, $A$                              &       &       & \multirow{5}{*}{\ref{sec:atom}}\\
        Atomic matrix element                   & $\mathcal{M}_\alpha$                  & $-$   & $+$   & \\
        Relativistic enhancement factor         & $\mathcal{R}$                         &       &       & \\
        Wavefunction $|s\rangle, |p_j\rangle$ coefficients	& $\epsilon_s, \epsilon_{p_j}$ &    &       & \\
        Effective quantum numbers               & $\nu_s, \nu_{p_j}$                    &       &       & \\
        \hline
        Mass density (crystal)					& $\rho$ 								&       &       & \multirow{13}{*}{\ref{sec:crystal}}\\
        Heavy-nucleon number density            & $n_N$                                 &       &       & \\
        Volume of unit cell                     & $V_c$                                 &       &       & \\
        Potential energy density (crystal)		& $U$									&       &       & \\
        Strain									& $S_{\alpha \beta}$		            & $+$   & $+$   & \\
        Stress									& $T_{\alpha \beta}$		            & $+$   & $+$   & \\
        Electric field							& $E_{\alpha}$				            & $-$   & $+$   & \\
        Electric displacement					& $D_{\alpha}$				            & $-$   & $+$   & \\
        Elastic stiffness (at constant $D$)		& $c^D_{\alpha \beta, \gamma \delta}$ 	& $+$   & $+$   & \\
        Impermittivity (at constant $S$)	    & $\beta_{\alpha \beta}^S$		        & $+$   & $+$   & \\
        Piezoelectric tensor 					& $h_{\alpha,\gamma \delta}$ 			& $-$   & $+$   & \\
        Piezoaxionic tensor						& $\xi_{\alpha, \gamma \delta}$			& $-$   & $+$   & \\
        Electroaxionic polarizability		    & $\zeta_{\alpha \beta}$ 				& $+$   & $+$   & \\
        \hline
        Impedance of circuit element $X$        & $Z_X$                                 &       &       & \multirow{5}{*}{\ref{sec:setup}} \\
        Inductance                              & $L_X$                                 &       &       & \\
        Capacitance                             & $C_X$                                 &       &       & \\
        Resistance                              & $R_X$                                 &       &       & \\
        Mutual inductance, coupling                       & $M_X, k_X$                                 &       &       & \\
        \hline
        Crystal length in direction $i$			& $\ell_i$								&       &       & \multirow{9}{*}{\ref{sec:signal}} \\
        Mechanical resonance frequency          & $f_0$                                 &       &       & \\
        Natural resonance frequency             & $f_\mathrm{nat}$                      &       &       & \\
        Mechanical displacement                 & $u_i$                                 &       &       & \\
        Sound speed (longitudinal)              & $v^D$                                 &       &       & \\
        Electromechanical coupling              & $k$                                   &       &       & \\
        Quality factor                          & $Q(\omega)$                           &       &       & \\
        Axion-induced voltage, current                   & $V_a$, $I_a$                                 &       &       & \\
        SQUID flux                              & $\Phi_\mathrm{SQ}$                    &       &       & \\
        \hline
        Noise spectral density (voltage)        & $S_V$                                 &       &       & \multirow{5}{*}{\ref{sec:backgrounds}} \\
        Loss angle of $X$							& $\delta_X$					& & & \\
        Temperature                             & $T$                                   &       &       & \\
        SQUID noise parameter               & $\eta_\mathrm{SQ}$                           &       &       & \\
        Cooling power, decay heat               & $\dot{Q}$                             &       &       & \\
        \hline
        Signal-to-noise ratio                   & $\mathrm{SNR}$                        &       &       & \multirow{4}{*}{\ref{sec:sensitivity}}\\
        Resonance frequency (circuit)     & $f_\mathrm{res}$                      &       &       & \\
        Shot time                               & $t_\mathrm{shot}$                     &       &       & \\
        Total integration time                  & $t_\mathrm{int}$                      &       &       & \\
 	\hline \hline
    \end{tabular}
    \caption{List of primary symbols appearing in the four subsections of both Secs.~\ref{sec:theory} and~\ref{sec:experiment}, their transformations ($\text{even}=+$, $\text{odd}=-$) under parity $\mathsf{P}$ and time-reversal $\mathsf{T}$ (where relevant), and the subsections where they first appear.}
    \label{key}
\end{table}

\section{Theory}\label{sec:theory}
In this section, which is split into four parts, we work out the theory of the piezoaxionic and electroaxionic effects. In Sec.~\ref{sec:axion}, we summarize some basic properties of the QCD axion and its fundamental coupling to gluons. Section~\ref{sec:nucleus} reviews the generation of $\mathsf{P}$- and $\mathsf{T}$-violating electromagnetic moments of nuclei due to this coupling, in particular the Schiff moment. Section~\ref{sec:atom} explains how a Schiff moment leads to energy shifts at the atomic level, with further details on atomic matrix elements provided in App.~\ref{app:matrixel}. Finally, in Sec.~\ref{sec:crystal}, a mapping of these atomic energy shifts to the piezoaxionic and electroaxionic crystal tensors is outlined. 

\subsection{Axion}\label{sec:axion}

The fundamental, defining interaction of the QCD axion $a(t,\vect{x})$ is its coupling to the QCD anomaly $\mathcal{G} \tilde{\mathcal{G}}$:
\begin{eqnarray}\label{eq:axion-gluon-coupling}
\mathcal{L} \supset \frac{\alpha_s^2}{8\pi}\left(\overline{\theta} + \frac{a}{f_a}\right) G^a_{\mu \nu} \widetilde{G}^{a\mu\nu}\equiv \frac{\alpha_s}{8 \pi}\bar{\theta}_a G^a_{\mu \nu} \widetilde{G}^{a\mu\nu}.
\end{eqnarray}
In the above Lagrangian $\mathcal{L}$, the static theta angle $\overline{\theta}$ parametrizes the static $\mathsf{CP}$ violation in the strong interactions, with strong coupling $\alpha_s$ to the canonically normalized gluon field strength $G^{a}_{\mu\nu}$ (with $\widetilde{G}^{a\mu \nu} \equiv \epsilon^{\mu \nu \rho \sigma} G^a_{\rho \sigma}/2$). The resulting non-perturbative potential for the axion field is minimized at $\overline{\theta}_a \simeq 0$, thus dynamically solving the strong $\mathsf{CP}$ problem in the vacuum~\cite{Wilczek:1977pj,Weinberg:1977ma,Peccei:1977hh}. Axion particles are effectively the excitations in this potential, and can constitute the DM in our Universe, with an rms $\overline{\theta}_a$ angle predicted to be~\cite{marsh2016axion}:
\begin{eqnarray}\label{eq:thetaQCDaxion}
\sqrt{\left \langle \overline{\theta}_a^2 \right \rangle }=\frac{\sqrt{2 \rho_a}}{m_a f_a } \approx 4\times 10^{-19} \sqrt{\frac{\rho_a}{\rho_\mathrm{DM}}},
\end{eqnarray}
where $\rho_a$ and $\rho_\mathrm{DM} \approx 0.4\,\mathrm{GeV}\,\mathrm{cm}^{-3}$ are the local cosmic-axion and (measured) DM energy densities~\cite[{\S}27]{Zyla:2020zbs}, respectively. The theta angle is odd under both parity and time-reversal, i.e.~$\overline{\theta}_a \to -\overline{\theta}_a$ under $\mathsf{P}$ and $\mathsf{T}$ separately.

The field ${\overline{\theta}_a}$ oscillates with a frequency $f = m_a/(2\pi)$ set by the axion mass $m_a$, which is related to the axion decay constant $f_a$ as~\cite{di2016qcd}:
\begin{align}
    m_a \approx \left(5.70 \pm 0.07\right) \times 10^{-10}\,\mathrm{eV} \left(\frac{10^{16}\,\mathrm{GeV}}{f_a}\right), \label{eq:mafa}
\end{align}
or in terms of frequency, $f = 0.138 \, \mathrm{MHz} \, ({10^{16}\,\mathrm{GeV}}/{f_a}).$
This inverse relation between the mass and the axion decay constant is why the typical amplitude of $\overline{\theta}_a$ of Eq.~\ref{eq:thetaQCDaxion} is independent of mass.
The coherence time $\tau_\mathrm{coh}$ of the axion field is determined by the inverse kinetic energy spread of the axion DM, given by:
\begin{align}
    \tau_\mathrm{coh} \simeq \frac{2\pi}{m_a} \frac{2}{v_0^2} = \frac{1}{f} \left(3.3 \times 10^6\right),\label{eq:taucoh}
\end{align}
where we have used the local Milky-Way DM velocity dispersion $v_0 \approx 235\,\mathrm{km/s}$~\cite[{\S}27]{Zyla:2020zbs}.

The mass-coupling relation of Eq.~\ref{eq:mafa} can be broken in principle. In practice, a modification to lower the $m_a f_a$ product of the QCD axion, which would increase the effective theta angle amplitude in Eq.~\ref{eq:thetaQCDaxion}, requires a fine tuning of both the first and second derivatives of the potential, to preserve the Peccei-Quinn solution to the strong $\mathsf{CP}$ problem and to partially cancel the QCD contributions to the axion mass, respectively. Even with significant model-building efforts, this part of parameter space is precarious, as finite-density and finite-temperature effects can undo the fine tuning in the vacuum, leading to wildly different in-medium minima in stellar systems~\cite{hook2018probing}. (We return to signatures and potential constraints from this effect in Sec.~\ref{sec:sensitivity}.) A significant increase in the $m_a f_a$ product is generically only fine-tuned in the first derivative of the potential, so that $\bar \theta \approx 0$ in Eq. (\ref{eq:axion-gluon-coupling}), but would lower the rms $\overline{\theta}_a$ angle and thus render the effects discussed in this work less significant. 

\subsection{Nucleus}\label{sec:nucleus}

In this section, we compute the $\overline{\theta}_a$-dependent Schiff moment of nuclei. The most basic $\mathsf{P}$- and $\mathsf{T}$-violating electromagnetic moment of a nucleus is its electric dipole moment (EDM). (An EDM is $\mathsf{P}$-odd and $\mathsf{T}$-even, but its alignment with spin, which itself is $\mathsf{P}$-even and $\mathsf{T}$-odd, is what violates $\mathsf{P}$ and $\mathsf{T}$ separately.)  Our ability to probe nuclear EDMs in atomic systems is, however, restricted by Schiff's screening theorem~\cite{schiff1963, Chupp2019,Pospelov:2005pr,Khriplovich1997ga,ENGEL201321}. In any system composed of nonrelativistic, pointlike, charged particles (e.g.~an atom or crystal), the bare EDMs of all particles are perfectly screened by a spatial rearrangement of the (monopole) electric charges. However, nuclei have a finite radius $R_0$, so the position of their EDMs need not coincide with their centers of charge. One can systematically take into account these finite-size effects through a nuclear multipole expansion: an EDM at dimension 1, a magnetic quadrupole moment (MQM) at dimension 2, an electric octupole moment (EOM) and a Schiff moment at dimension 3, etc. The EDM is screened, the MQM only has effects in magnetic materials, and effects on the electronic wavefunction from the EOM are suppressed by the angular-momentum barrier near the nucleus~\cite{Khriplovich1997ga}. (Moments at dimension 4 and higher are completely negligible.)  

Hence our focus on the Schiff moment, which interacts with electrons via the Hamiltonian:
\begin{equation}
H_\mathsf{S} =  4 \pi e \vect{\mathsf{S}}  \cdot \vect{\nabla} \delta(\vect{r}), 
\label{eq:schiff potential}
\end{equation}
where $\vect{\nabla} \delta(\vect{r})$ is to be evaluated on the electron wavefunction. Like an EDM, a Schiff moment is $\mathsf{P}$-odd and $\mathsf{T}$-even, but its alignment with the nuclear spin violates both $\mathsf{P}$ and $\mathsf{T}$ invariance. Note that the effective operator on electrons is $\mathsf{P}$-odd and $\mathsf{T}$-even. A Schiff moment~\cite{ENGEL201321}
\begin{eqnarray}
\Sch = \Sch_\text{EDM} + \Sch_\text{ch}
\end{eqnarray}
can arise from EDMs of the constituent nucleons
\begin{align}
\Sch_\text{EDM} = \sum_{j = p, n} \Bigg \lbrace & \frac{1}{6} e \left[ \vect{r}_j (\vect{r}_j\cdot \vect{d}_j)-\frac{1}{3}\braket{r^2}_\text{ch}\vect{d}_j \right] \label{eq:SchiffEDM} \\
& + \frac{1}{6} e \left( r_j^2 \vect{r}_j -\braket{r^2}_\text{ch}\vect{r}_j \right) \Bigg \rbrace \nonumber
\end{align}
and from an asymmetrical distribution of charge (i.e.~protons) within the nucleus
\begin{align}
    \Sch_\text{ch}=\sum_p \frac{1}{10} e \left( r_p^2 \vect{r}_p-\frac{5}{3}\braket{r^2}_\text{ch}\vect{r}_p \right). \label{eq:Schiffch}
\end{align}
In Eq.~\ref{eq:SchiffEDM}, the sum over $j$ runs over all nucleons (protons $p$ and neutrons $n$), while the sum in Eq.~\ref{eq:Schiffch} runs over all protons. Above, we have also used shorthand for the mean squared radius of the nuclear charge distribution (with density $\rho_\text{ch}$):
\begin{equation}
     \braket{r^2}_\text{ch} \equiv \frac{1}{Z}\int \dd ^3 \vect{r} \, r^2 \rho_\text{ch}(\vect{r}),
\label{eq:rq}
\end{equation} 
with $Z$ the total nuclear charge. The \emph{magnitude} of the Schiff moment is defined as $\mathsf{S} \equiv \braket{\mathsf{S}_z}_{M = I}$, i.e.~the magnitude of the expectation value of $\Sch$ in the $+z$-direction, for a fully polarized nuclear spin along the $+z$-direction ($M = I$).

As we derive below, the effects from the Schiff moment are largest in high-$Z$ nuclei, because of their larger size and because of relativistic enhancement effects of the electron density for heavy nuclei (see Sec.~\ref{sec:atom} for this second effect). In Sec.~\ref{sec:SchiffEDM}, we evaluate the size of the Schiff moments from bare nucleon EDMs through Eq.~\ref{eq:SchiffEDM}. In Sec.~\ref{sec:PTforces}, we review the $\mathsf{PT}$-violating nuclear forces that arise at finite $\overline{\theta}_a$, and then evaluate the resulting Schiff moments via Eq.~\ref{eq:Schiffch} in non-deformed nuclei in Sec.~\ref{sec:nondeformed}, and in octupole-deformed nuclei in Sec.~\ref{sec:deformed}.

\subsubsection{Schiff moment from the EDM of a valence nucleon} \label{sec:SchiffEDM}

A theta term, from a static $\overline{\theta}$ or from an oscillating $\overline{\theta}_a$ in a QCD axion DM background, induces a typical nucleon EDM of order~\cite{Crewther:1979pi, Pospelov:2005pr}:
\begin{align}
d_{n} \approx 10^{-3}\, \overline{\theta}_a \,\text{e fm}. \label{eq:nucleonEDM}
\end{align}
This nucleon EDM in turn produces a nuclear Schiff moment in nuclei with odd atomic number $A$. For an odd-$A$ nucleus with a static, spherically symmetric core of radius $R_0 \simeq r_0 A^{1/3}$ with $r_0 \approx 1.2\,\mathrm{fm}$, the valence nucleon EDM leads to a Schiff moment magnitude~\cite{Khriplovich1997ga}:
\begin{align}
    \mathsf{S}_{\text{EDM}} &= \frac{1}{10} d_n R_0^2 \frac{2}{5}\frac{(K+1)}{(I+1)} \label{eq:SchiffEDMestimate}\\
    &\approx 2 \times 10^{-3} \, \overline{\theta}_a \,e\,\text{fm}^3 \frac{(K+1)}{(I+1)} \left(\frac{A}{230}\right)^{2/3}, \nonumber
\end{align}
with $K\equiv(l-I)(2I-1)$, $I$ the nuclear spin, and $l$ the valence nucleon's orbital angular momentum. 
The nuclear core may exhibit significant polarizability and deformations from sphericity, in which case the estimate of Eq.~\ref{eq:SchiffEDMestimate} will receive large corrections. However, in those cases, the $\mathsf{S}_{\text{EDM}}$ contribution to the Schiff moment will be subdominant to the asymmetric charge distribution of the nucleus, to which we turn next. 

\subsubsection{\texorpdfstring{$\mathsf{P-}$}~and \texorpdfstring{$\mathsf{T-}$}~violating forces} \label{sec:PTforces}

Forces between nucleons that violate parity and time-reversal symmetries can cause an asymmetric (parity-odd) charge distribution of the nucleus that is aligned with the nuclear spin, thus generating a Schiff moment $\mathsf{S}_\mathrm{ch}$ via Eq.~\ref{eq:Schiffch}~\cite{sushkov1984,Haxton_Henley_1983,griffiths1991}. Since this is a tree-level effect from light-meson exchange between nucleons, this contribution to the Schiff moment will typically dominate over the loop-level effect from Eqs.~\ref{eq:nucleonEDM} and~\ref{eq:SchiffEDMestimate}. To leading order, these $\mathsf{P}$- and $\mathsf{T}$-violating forces are mediated by pion exchange; in the heavy-pion limit ($m_\pi R_0 \gg 1$) and nonrelativistic approximation, they are given by the effective interaction potential:
\begin{align}
    V_{\mathsf{P}\mathsf{T}} =\frac{1}{2 m_N m_\pi^2} (\eta_{ab}\bm{\sigma_a}-\eta_{ba}\bm{\sigma_b})\cdot\bm{\nabla}\delta(\bm{r}_a-\bm{r}_b).
\label{eq:V_PT}
\end{align}
where $m_N \simeq m_n \simeq m_p$ is the nucleon mass, the Pauli matrix $\vect{\sigma}$ acts on the spin Hilbert space of the nucleons $a$ and $b$, and $\vect{r}$ on the position Hilbert space. The effective couplings among protons $p$ and neutrons $n$ are:
\begin{alignat}{3}
    \eta_{pp} &= -\eta_{np} &&= g_s(g_0+g_1),\\
    \eta_{nn} &= -\eta_{pn} &&= g_s(g_1-g_0).
\end{alignat}
The coupling $g_s$ is the standard $\mathsf{P}$- and $\mathsf{T}$-conserving pion-nucleon coupling, while $g_0$, and $g_1$ are the $\mathsf{P}$- and $\mathsf{T}$-violating isoscalar and isovector couplings, respectively, sourced by the $\overline{\theta}_a$ term. We assume the numerical values~\cite{deVries:2020iea}:
\begin{align}
    g_s &\approx -13.45, \\
    g_0 &\approx (15.5 \pm 2.6) \times 10^{-3} \, \overline{\theta}_a,\\
    g_1 &\approx -0.2 g_0.
\end{align}

We neglect the numerically smaller isotensor pion-nucleon coupling $g_2$. The contributions to $V_{\mathsf{P}\mathsf{T}}$ from exchange of $\rho$ and $\omega$ mesons have been argued by Refs.~\cite{herczeg_1995,Towner:1994qe} to be smaller than the pion-exchange contributions from Eq.~\ref{eq:V_PT}, by the numerical factor $0.3(m_\pi^2/m_{\rho,\omega}^2)$. 

\subsubsection{Non-deformed nuclei} \label{sec:nondeformed}
The $\mathsf{P}$- and $\mathsf{T}$-odd internucleon potential of Eq.~\ref{eq:V_PT} provides the symmetry breaking needed to create a $\mathsf{P}$-odd charge distribution of a spin-polarized nucleus. Since the values of $\overline{\theta}_a$ of interest are tiny (cfr.~Eq.~\ref{eq:thetaQCDaxion}), we will evaluate the Schiff moment in perturbation theory:
\begin{align}
    \braket{\vect{\mathsf{S}}_\mathrm{ch}} = \sum_n \frac{\braket{0| V_{\mathsf{P}\mathsf{T}} | n}\braket{n| \vect{\mathsf{S}}_\mathrm{ch} | 0}}{E_0-E_n}+\mathrm{h.c.},
\label{eq:Schiff_perturbation}
\end{align}
where the sum is over nuclear states $n$ with energies $E_n$; $n=0$ labels the ground state.

In the nuclear shell model, where the core is treated as a static source, only a valence proton can generate a nonzero Schiff moment. The effective mean-field interaction potential of the valence nucleon (with label $a$) stemming from Eq.~\ref{eq:V_PT} becomes:
\begin{align}
    V_{\mathsf{P}\mathsf{T}}&=\frac{1}{2 m_N m_\pi^2}\eta_{a}\vect{\sigma}_a \cdot \vect{\nabla}_a\rho(\vect{r}_a), \label{eq:H_PT}
\\
    \eta_a & \equiv \eta_{ap} \frac{Z}{A} +\eta_{an} \frac{A-Z}{A}, \label{eq:etaa}
\end{align}
with $\rho$ the number density of core neutrons and protons normalized as $\int \dd^3r\, \rho (\vect{r}) = A$. (In the above formulae, we ignore corrections of order $1/A$ and $1/Z$.) The contribution to the Schiff moment from a valence proton in the nuclear shell model is~\cite{sushkov1984}:
\begin{align}
    \mathsf{S}_\mathrm{ch} \simeq \frac{e}{10}\frac{\eta_a}{2m_N m_\pi^2}\frac{\rho_0}{U_0} R_0^2 \simeq \frac{3 e \eta_a}{4 \pi^2 m_N} R_0^2, \label{eq:Schiff_nucshell}
\end{align}
with $\rho_0 \simeq 3/(4\pi r_0^3)$ the nucleon number density, and $U_0 \simeq 3 \pi/(60 m_\pi^2 r_0^3)$ the spherically symmetric potential of the nuclear core.

However, excitations of \emph{core protons}, which are dynamics not captured within the nuclear shell model, by either a valence proton or neutron, can yield Schiff moment contributions of similar size as the expression in Eq.~\ref{eq:Schiff_nucshell}~\cite{Flambaum:1985gv}. This means that $\mathsf{P}$- and $\mathsf{T}$-odd forces can generate Schiff moments in nuclei with either an odd number of protons or an odd number neutrons. Numerically, the effect in Eq.~\ref{eq:Schiff_nucshell} is typically much larger than that of Eq.~\ref{eq:SchiffEDMestimate}. For example, one finds:
\begin{align}
    \mathsf{S}_\mathrm{ch}\left(\ce{^{181}_{73}Ta}\right) \simeq 0.17 \, \overline{\theta}_a\, e \, \text{fm}^3 \label{eq:Schiff_Ta}
\end{align}
for the stable nucleus of $\ce{^{181}_{73}Ta}$, which has one valence proton. Estimates for other stable isotopes of heavy nuclei can be found e.g.~in Ref.~\cite{ENGEL201321}.

\subsubsection{Octupole-deformed nuclei} \label{sec:deformed}
A class of nuclei with strongly enhanced $\mathsf{P}$- and $\mathsf{T}$-violating moments are those possessing \emph{permanent octupole deformations} or \emph{soft octupole modes}.  (For reviews on deformed nuclei, see Refs.~\cite{Butler:2016rmu, Krane:359790, butler2020pear, cao2020landscape}.) Such nuclei typically have anomalously large values for certain transition matrix elements, paired with small energy splittings between opposite-parity nuclear levels. The origin of these two signatures and their role in collective enhancements of the nuclear Schiff moment will be summarized here, following Refs.~\cite{Spevak:1996tu, Engel:1999np}.

We first discuss permanent octupole deformation of heavy nuclei. Analogously to the Born-Oppenheimer approximation in molecules, where one factorizes rotational motion of a ``rigid'' shape from the internal excitations (vibration, electronic excitations, etc.), we can consider the shape of the nucleus in the intrinsic (body-fixed) frame of the nucleus. Assuming the nuclear core to be axially symmetric and of constant density (equal for protons and neutrons), the core shape is described by the location of its surface $R$, expanded into the following multipoles $l$:
\begin{align}
    R = R_0\left(1+\sum_{l=1} \beta_l Y_{l0}\right),
\label{eq:deformed_surface}
\end{align}
where the $\beta_i$ parameters characterize the strength of the dipole, quadrupole, octupole, etc. deformations. The dipole deformation $\beta_1$ is fixed such that the center of mass/charge is at the origin. In the intrinsic frame, and with the above assumptions, the Schiff moment can then be straightforwardly calculated using Eq.~\ref{eq:Schiffch}:
\begin{equation}
    \mathsf{S}_\mathrm{ch}^\mathrm{(int)} = Z e R_0^3 \frac{3}{20 \pi}\sum_{l=2}\frac{(l+1)\beta_l\beta_{l+1}}{\sqrt{(2l+1)(2l+3)}},
\end{equation}
with the main piece typically coming from the first term $\propto \beta_2 \beta_3$.

If a deformed nucleus is reflection asymmetric in its intrinsic frame, then in the laboratory frame, its ground state wavefunction will be composed of a parity doublet:
\begin{equation}
    \Psi^\pm = \frac{1}{\sqrt{2}}(\ket{ IMK } \pm \ket{ IM-K }).
\label{eq:parity_doublet}
\end{equation}
Here $I$ is the nuclear spin, $M$ is the quantum number of $I_z$, and $K$ that of the operator $\vect{I}\cdot\vect{n}$, with $\vect{n}$ the nuclear axis. The wavefunctions $\Psi^\pm$ are good parity states since $\braket{\Psi^\pm|\vect{n}|\Psi^\pm}=0$, i.e.~there is no average orientation of the nuclear axis, and $\mathsf{P}$ and $\mathsf{T}$ are preserved. Turning around the argument, a tell-tale signature of static octupole deformation of a nucleus is the presence of a low-lying opposite-parity level with the same angular momentum as the ground state. (In reality, the members of the parity doublet are not exactly degenerate due to Coriolis forces and other effects; these are analogous to the rovibrational and vibronic couplings that signal corrections to the Born-Oppenheimer approximation for molecular states.)

Interactions that violate both $\mathsf{P}$ and $\mathsf{T}$ can mix these opposite-parity states of a spin-polarized nucleus, and partially align the nuclear axis with the nuclear spin, leading to a collective enhancement of the Schiff moment. The perturbed wavefunctions under the interaction Hamiltonian of Eq.~\ref{eq:H_PT} are:
\begin{equation}
    \widetilde{\Psi}^+ = \Psi^+ + \widetilde{\alpha} \Psi^-, \quad \widetilde{\Psi}^- = \Psi^- - \widetilde{\alpha} \Psi^+,
\end{equation}
where
\begin{equation}
    \widetilde{\alpha} = \frac{\braket{\Psi^-| V_{\mathsf{P}\mathsf{T}}| \Psi^+}}{\Delta E_\pm}\approx 0.5 \,\overline{\theta}_a \frac{\eta_a}{0.1}\frac{50 \, \text{keV}}{\Delta E_\pm} \frac{\beta_3}{0.1} \left(\frac{230}{A}\right)^{1/3},
 \end{equation}
with $\Delta E_\pm \equiv E^+-E^-$ the energy splitting of the doublet and $\eta_a$ the effective coupling of the valence nucleon as in Eq.~\ref{eq:etaa}. Within this framework, we then finally have $\braket{\widetilde{\Psi}^\pm | n_z | \widetilde{\Psi}^\pm } = 2 \widetilde{\alpha} K M / [(I+1) I]$, and since the ground-state wavefunction typically has $K=I$~\cite{Spevak:1996tu}, a laboratory-frame Schiff moment of
\begin{align}
    \mathsf{S}_\mathrm{ch} &= 2 \widetilde{\alpha} \frac{I}{I+1} \mathsf{S}_\mathrm{ch}^\mathrm{(int)} \label{eq:schiffoctupole} \\
    &\approx 5 \,\overline{\theta}_a\, e\,\mathrm{fm}^3 \, \frac{\eta_a}{0.1} \frac{50 \, \text{keV}}{\Delta E_\pm} \frac{\beta_2}{0.12}\left(\frac{\beta_3}{0.1}\right)^2\frac{Z}{88}\left(\frac{A}{230}\right)^{2/3}. \nonumber
\end{align}
Contributions from higher-order deformations of the nucleus are usually subdominant.
We observe that the collective ($Z$-enhanced) effect of static octupole deformation can lead to very large Schiff moments, compared to the contributions from valence nucleons in non-deformed nuclei (Eqs.~\ref{eq:Schiff_nucshell} and~\ref{eq:Schiff_Ta}) and from bare EDMs of the valence nucleons (Eq.~\ref{eq:SchiffEDMestimate}). Tabulations of the parameters $\beta_2$, $\beta_3$, and $\Delta E_\pm$ can be found in Ref.~\cite{MOLLER20161} for the ground states of nearly all nuclei, though estimates with other nuclear structure models often give different results~\cite{Ebata:2017npw,nazarewicz1984analysis}.

The class of nuclei with collectively enhanced Schiff moments extends beyond those with static octupole deformations, and includes also those with \emph{soft octupole modes}. Ref.~\cite{Engel:1999np} showed that it is sufficient to have a significant octupole-deformation-\emph{squared}, i.e.~$\braket{\beta_3^2} \neq 0$, even if $\braket{\beta_3} = 0$, for a nucleus to have a $Z$-enhanced Schiff moment. This dynamical octupole deformation can be thought of as a collective nuclear octupole vibration with angular frequency $\Delta E_\pm$, and corresponding zero-point amplitude-squared inversely proportional to this frequency, $\braket{\beta_3^2} \propto 1/\Delta E_\pm$, as for any quantum harmonic oscillator. 

An ideal octupole-enhanced candidate nuclear isotope (with either static or dynamical deformation) should thus have an opposite-parity level with a small energy gap above the ground state. For the experimental concept proposed in this paper, the isotope needs to be (meta)stable (cfr.~the discussion on heating from radioactivity in Sec.~\ref{sec:backgrounds}). Energy splittings and half-lives are tabulated in Ref.~\cite{nudat2}. Three potentially suitable isotopes were suggested in Ref.~\cite{Flambaum2020}, together with estimates of their Schiff moments:
\begin{alignat}{5}
 &\mathsf{S}_\mathrm{ch}\left(\ce{^{153}_{63}Eu}\right) &&\approx  3.7\, \overline{\theta}_a\, e \,\text{fm}^3,~ && \\
 &\mathsf{S}_\mathrm{ch}\left(\ce{^{235}_{92} U}\right) &&\approx 3.0 \,\overline{\theta}_a \, e \,\text{fm}^3,~ && \tau_{1/2} \approx 7 \times 10^8 \, \mathrm{yr}, \\
 &\mathsf{S}_\mathrm{ch}\left(\ce{^{237}_{93}Np}\right) &&\approx 6.0\,\overline{\theta}_a\, e \,\text{fm}^3,~ &&\tau_{1/2} \approx 2 \times 10^6 \, \mathrm{yr}.
\end{alignat}
In addition, the isotopes $\ce{^{161}_{66}Dy}$ and $\ce{^{155}_{64}Gd}$ are known to have very small energy gaps between the ground state and the excited state of opposite parity~\cite{nudat2}, which suggests a possible octupole enhancement. According to Ref.~\cite{MOLLER20161}, they are not statically octupole deformed, but they could still exhibit significant dynamical octupole deformation. We therefore estimate the Schiff moment from soft octupole vibrations using Eq.~\ref{eq:schiffoctupole} as suggested in Ref.~\cite{Engel:1999np}, with the squared octupole deformation estimated using the collective $B(E3)$ octupole transition probability for neighbouring even-even nuclei found in e.g.~Ref.~\cite{spear2002}:
\begin{equation}
  B(E3)_{0^+\to 3^-} = \left(\frac{3}{4\pi}\right)^2 \left(ZeR_0^3\right)^2 \braket{\beta_3^2}. \label{eq:BE3}
\end{equation}
Data from Ref.~\cite{nudat2} with Eqs.~\ref{eq:schiffoctupole} and~\ref{eq:BE3} then yields our estimates:
\begin{alignat}{3}
  &\mathsf{S}_\mathrm{ch}\left(\ce{^{161}_{66}Dy}\right) &&\approx 4 \, \overline{\theta}_a\, e \,\text{fm}^3,\\
  &\mathsf{S}_\mathrm{ch}\left(\ce{^{155}_{64}Gd}\right) &&\approx 1 \, \overline{\theta}_a\, e \,\text{fm}^3.
\end{alignat}
Despite these encouragingly large fiducial values, the accuracy of the predictions in this section is under poor control, as different nuclear structure models suggest widely different values~\cite{butler2020pear, cao2020landscape}. More theoretical support is needed to identify suitable candidate isotopes for a large-scale experiment, and to provide accurate and precise determinations of the predicted Schiff moments in the presence of a nonzero $\overline{\theta}_a$ parameter.

\subsection{Atom}\label{sec:atom}
In Eq.~\ref{eq:schiff potential} of Sec.~\ref{sec:nucleus}, we already wrote down the parity-odd electrostatic potential for electrons produced by a nuclear Schiff moment. This potential can mix opposite-parity electronic states, or equivalently, lead to an atomic energy shift if the electronic wavefunction already breaks parity (which is the case inside a piezoelectric crystal).

The energy shift of an atom with a polarized nuclear spin and Schiff moment, from the perturbing Hamiltonian in Eq.~\ref{eq:schiff potential}, is:
\begin{align}
\langle H_{\mathsf{S}} \rangle &= - 4 \pi e \mathsf{S} \sum_{j,m_j,k}\hat{I}_k \epsilon_s \epsilon^*_{p_{j,m_j}} \mathcal{M}_{j,m_j,k}+ \text{c.c.},\\
&\equiv - 4\pi e \mathsf{S} \sum_k \mathcal{M}_k \hat{I}_k;
\label{eq:<deltaH>}
\end{align}
where the quantum numbers $j$ and $m_j$ are those of the total angular momentum and its projection on the $z$-axis, respectively, the index $k$ denotes spatial direction, $\hat{I}_k$ is the direction of the nuclear spin (normalized such that $|\hat{\vect{I}}| = 1$ corresponds to a fully polarized nuclear spin state, and $|\hat{\vect{I}}| = 0$ an unpolarized state), and $\mathcal{M}_{j,m_j,k} =  \bra{s^0} \hat{r}_k \, \partial_r \delta^3(\vect{r}) \ket{ p^0_{j, m_j}}$ is an atomic matrix element. The coefficients $\epsilon_{s}$ and $\epsilon_{p_{j, m_j}}$ parametrize the admixture of atomic $s$ and $p_{j, m_j}$ valence electron states, i.e.~$\ket{\psi_{\text{el}}} = \epsilon_s\ket{s^0} + \sum_{j, m_j} \epsilon_{p_{j, m_j}} \ket{p^0_{j, m_j}}$, and characterize the breaking of parity symmetry by the crystal potential. (This treatment of the atom's wavefuction in a crystal is only possible in the tight-binding approximation, which holds to leading order for the insulating crystals considered in this paper.) The admixture of opposite-parity states in the atom's ground state in the crystal allows for a piezoaxionic effect \emph{linear} in the Schiff moment, without the need of other parity-breaking sources such as the application of an external electric field. We expand on this procedure in App.~\ref{app:matrixel}.

\begin{figure}
    \includegraphics[width=0.48\textwidth]{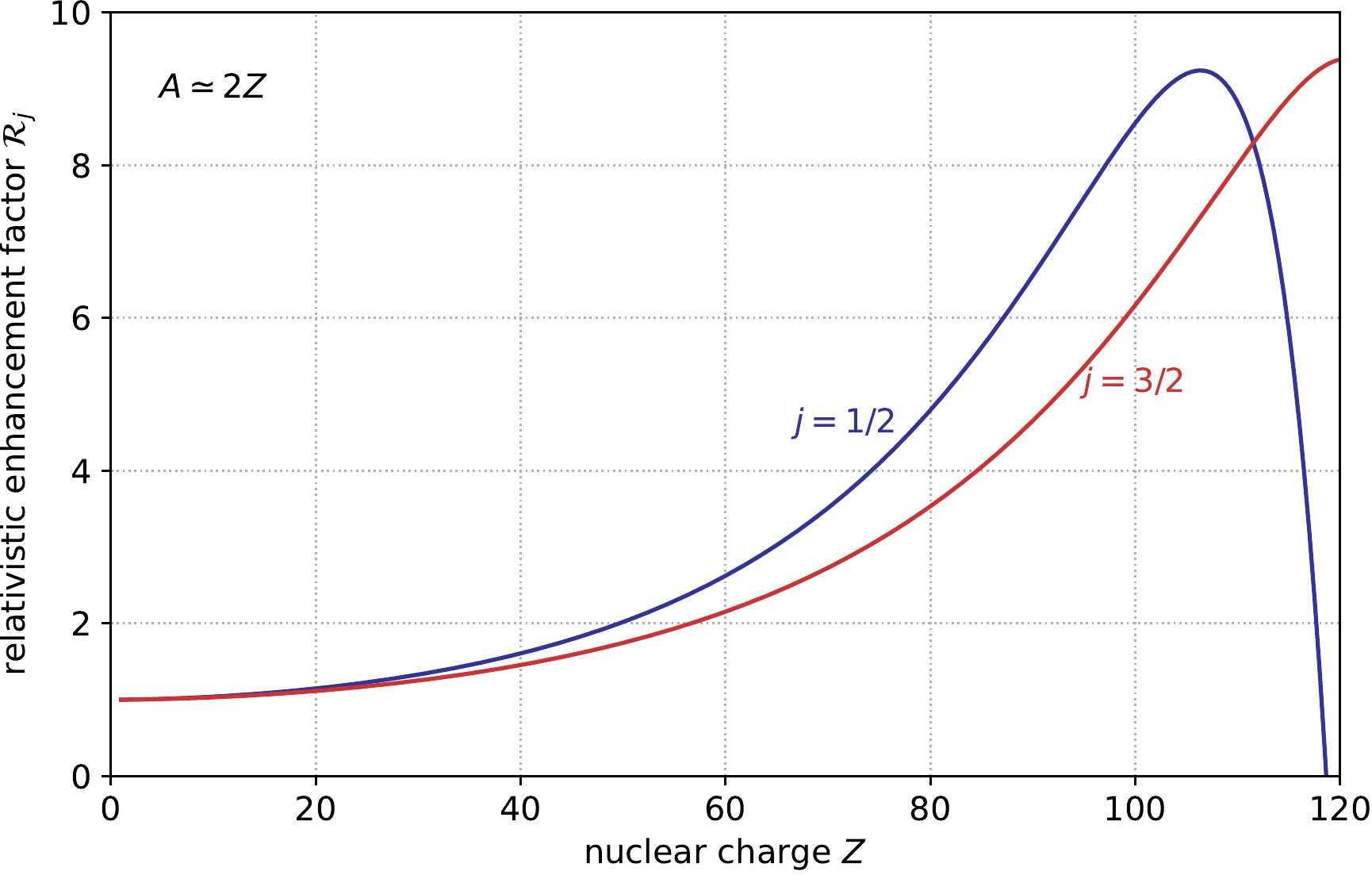}
    \caption{The relativistic enhancement factor $\mathcal{R}_j$ of Eqs.~\ref{eq:R12} and~\ref{eq:R32} as a function of the proton number $Z$, assuming an atomic number of $A \approx 2Z$, for two values of the total electron angular momentum $j = 1/2, 3/2$. \nblink{supporting_plots_eta20.ipynb}}
    \label{fig:rel_enhancement}
\end{figure}

The energy shift of Eq.~\ref{eq:<deltaH>} can be reliably calculated in perturbation theory up to a wavefunction normalization constant, as the Schiff interaction of Eq.~\ref{eq:schiff potential} is a short-distance effect where screening effects are unimportant. Armed with only a few inputs, namely $E_s \equiv \alpha^2 m_e / 2 \nu_s$ and $E_{p_j} \equiv \alpha^2 m_e / 2 \nu_{p_j}$ with effective quantum numbers $\nu_s$ and $\nu_j$, the normalization of the radial part of the wavefunctions can be fixed by the matching procedure of Ref.~\cite[Ch.~8]{Khriplovich1997ga}. This leads to a matrix element of the form:
\begin{equation}
   \mathcal{M}_{{j, m_j},k} =  \braket{\Omega_s\lvert {\hat{r}_k}\rvert \Omega_{p,{j, m_j}}} \frac{Z^2}{a_0^4(\nu_s\nu_{p_j})^{3/2}}\mathcal{R}_j
\label{eq:renormalisedM}
\end{equation}
where $a_0 \equiv 1/\alpha m_e$ is the Bohr radius, $\Omega$ are the spinor spherical harmonics as in Ref.~\cite[{\S}24]{lifshitz1974relativistic}, and their angular matrix elements in Eq.~\ref{eq:renormalisedM} are $\mathcal{O}(1)$ and evaluated in App.~\ref{app:matrixel}. The relativistic enhancement factors $\mathcal{R}_j$ are defined as:
\begin{align}
        \mathcal{R}_{\half} &= \frac{3 \gamma_{\half}(2\gamma_{\half}-1)}{2\gamma_{\half}+1}\frac{4}{\Gamma^2(2\gamma_{\half}+1)}\left( \frac{a_0}{2ZR_0}\right)^{2-2\gamma_{\half}} \label{eq:R12}
        \\
        \mathcal{R}_{\threehalf} &= \frac{6 \left[ (\gamma_{\half}+1)(\gamma_{\threehalf}+2)+Z^2\alpha^2\right]}{\Gamma(2\gamma_{\half}+1)\Gamma(2\gamma_{\threehalf}+1)},  \label{eq:R32} \\
        &\phantom{=} \times \left(\gamma_{\half}+\gamma_{\threehalf}-2\right) \left( \frac{a_0}{2ZR_0}\right)^{3-\gamma_{\half}-\gamma_{\threehalf}}, \nonumber
        \\ \gamma_j &= \sqrt{\left(j+\half\right)^2-Z^2\alpha^2}.
\end{align}
These enhancement factors are defined such that they go to unity in the limit $Z\alpha\to0$, and evaluate to about 3--8 for nuclei of interest, with $60 \lesssim Z\lesssim 93$, see Fig.~\ref{fig:rel_enhancement}.

A numerical estimate of the energy shift yields:
\begin{align}\label{eq:deltaH-num}
    \langle H_{\mathsf{S}} \rangle &\approx 2.0 \times 10^{-24}\,\mathrm{eV}\\
    &\phantom{\approx} \times \left(\frac{Z}{80}\right)^2 \frac{\mathcal{R} \left( \nu_s \, \nu_p\right)^{-\threehalf}}{10}\frac{\mathsf{S}}{\overline{\theta}_a\,e\,\text{fm}^3} \frac{\epsilon_s \epsilon_p \hat{I}_k \braket{\Omega_s|\hat{r}_k|\Omega_p}}{1}, \nonumber
\end{align}
where we assume $\overline{\theta}_a$ is given by the rms amplitude of Eq.~\ref{eq:thetaQCDaxion}, i.e.~that the QCD axion makes up all of the DM.

\subsection{Crystal}\label{sec:crystal}
Up to this point, we have reviewed how QCD axion DM gives rise to an oscillatory $\overline{\theta}_a$ angle (Sec.~\ref{sec:axion}), nuclear Schiff moments (Sec.~\ref{sec:nucleus}), and atomic energy shifts (Sec.~\ref{sec:atom}). In this section, we will show how the latter translates in effective stresses and electric displacements in macroscopic crystals.

The internal energy density $U$ of the crystal can be expanded to quadratic order around its equilibrium as:
\begin{align}
U =  \frac{\vect{S}^\intercal \vect{c}^D \vect{S}}{2} - \vect{D}^\intercal \vect{h} \vect{S} + \frac{\vect{D}^\intercal \vect{\beta}^S \vect{D}}{2} - \vect{S}^\intercal \vect{\xi} \hat{\vect{I}}  \overline{\theta}_a  - \vect{D}^\intercal \vect{\zeta} \hat{\vect{I}}  \overline{\theta}_a .
\label{eq:int-energy}
\end{align}
Here the independent variables are the strain 2-tensor $\vect{S}$, the electric displacement vector $\vect{D}$, the nuclear spin polarization direction $\hat{\vect{I}}$  (normalized so that saturated spin polarization corresponds to $|\hat{\vect{I}}| = 1$), and the axionic theta angle $\overline{\theta}_a$.  The proportionality constants of the first three terms are given by the elastic stiffness 4-tensor (at constant electric displacement) $\vect{c}^D$, piezoelectric 3-tensor $\vect{h}$, and dielectric impermittivity 2-tensor (at constant strain) $\vect{\beta}^S$. In the last two terms, we introduce new interactions at linear order in the nuclear Schiff moment direction, whose proportionality constants we will refer to as the ``piezoaxionic'' 3-tensor $\vect{\xi}$  and ``electroaxionic" 2-tensor $\vect{\zeta}$. In absence of these two nuclear-spin-induced contributions, the internal energy density above reduces to the usual expression for a piezoelectric crystal, which can be found e.g.~in Ref.~\cite[Ch.~5]{tiersten1969linear}. Superscripts $^\intercal$ denote transposed quantities. We assume all quantities are given at (or near) zero temperature. We have also integrated out short-wavelength fluctuations, such as individual atomic displacements, a more detailed treatment of which is given in App.~\ref{app:long}; all quantities in Eq.~\ref{eq:int-energy} should be understood to be (nearly) homogeneous.

The constitutive equations for the stress $\vect{T}$ and electric field $\vect{E}$ can be written as first derivatives of the internal energy density of Eq.~\ref{eq:int-energy}~\cite{tiersten1969linear},
\begin{alignat}{9}
    \vect{T} &= \frac{\partial U}{\partial \vect{S}}  &&=  +\vect{c}^D\vect{S} &&- \vect{h}^\intercal\vect{D} &&- \vect{\xi}\hat{\vect{I}} \overline{\theta}_a \label{eq:Tstress}, \\
    \vect{E} &=  \frac{\partial U}{\partial \vect{D}} &&= - \vect{h} \vect{S}  &&+ \vect{\beta}^S \vect{D}    &&- \vect{\zeta} \hat{\vect{I}} \overline{\theta}_a.\label{eq:Eelecfield}
\end{alignat}
Equations~\ref{eq:Tstress} and~\ref{eq:Eelecfield} reveal that an axion DM background in the presence of nuclear spin polarization manifests itself both as a stress across the crystal due to $\vect{\xi}$ and an electric field due to $\vect{\zeta}$.

For the subsequent discussion of our proposed experimental setup and sensitivity, it is instructive to introduce Voigt notation. In this notation, the independent components of $3\times3$ symmetric tensors such as the strain $S_{\alpha \beta}$ and stress $T_{\alpha \beta}$ are reduced to 6 dimensional ``vectors'' $S_i$ and $T_i$, and the indices of their proportionality constants are reduced similarly (see App.~\ref{app:long} for further details). The above equations can thus be written as:
\begin{alignat}{5}
    T_n &= +c^D_{nk}S_k     &&- h_{nk}^\intercal D_k       &&- \xi_{nk}\hat{I}_k \overline{\theta}_a, \label{eq:TVoigt}\\
    E_n &= - h_{nk}S_k      &&+\beta^S_{nk}D_k  &&- \zeta_{nk}\hat{I}_k \overline{\theta}_a; \label{eq:EVoigt}
\end{alignat}
with Einstein summation convention on repeated indices. 

The piezoaxionic tensor in Eq.~\ref{eq:int-energy} can be computed from the atomic matrix element $\mathcal{M}$ of Sec.~\ref{sec:atom} through
\begin{align}
\xi_{nk} &= \frac{\partial}{\partial \overline{\theta}} \sum_{t=1}^{N_\mathsf{S}} 4 \pi e \mathsf{S} \frac{\partial}{\partial S_n}  \sum_{j,m_j} \left[\frac{\epsilon_s \epsilon^*_{p_{j,m_j}} \mathcal{M}_{j,m_j,k}}{V_c} + \cc \right]_{(t)} \nonumber \\
&\equiv \widetilde{\xi}_{nk} \frac{4 \pi e  N_\mathsf{S}}{V_c} \frac{\dd{\mathsf{S}}}{\dd \overline{\theta}} \mathcal{M}_k, 
\label{eq:xidef}
\end{align}
which follows directly from Eq.~\ref{eq:<deltaH>}, and where the subscript $t = 1, \dots, N_\mathsf{S}$ runs over the spin-polarized nuclei (which are all assumed to have the same Schiff moment $\mathsf{S}$) in the unit cell with volume $V_c$. The electroaxionic tensor is given by a similar formula:
\begin{align}
\zeta_{nk} &=  \frac{\partial}{\partial \overline{\theta}}  \sum_{t=1}^{N_\mathsf{S}} 4 \pi e \mathsf{S} \frac{\partial}{\partial D_n}  \sum_{j,m_j} \left[\frac{\epsilon_s \epsilon^*_{p_{j,m_j}} \mathcal{M}_{j,m_j,k}}{V_c} + \cc \right]_{(t)} \nonumber \\
&\equiv \widetilde{\zeta}_{nk} \frac{4 \pi e N_\mathsf{S}}{V_c}\frac{e a_0^2}{\alpha}  \frac{\dd{\mathsf{S}}}{\dd \overline{\theta}}  \mathcal{M}_k.
\label{eq:zetadef}
\end{align}
In Eqs.~\ref{eq:xidef} and \ref{eq:zetadef}, we have defined the dimensionless quantities $\widetilde{\xi}$ and $\widetilde{\zeta}$, respectively, and are the only factors through which the electron wavefunction properties of the atoms in the crystal enter. 

A density functional theory (DFT) calculation of $\widetilde{\xi}$ and $\widetilde{\zeta}$ for materials of interest is left to future work. Here, we will content ourselves with naive dimensional analysis (NDA) estimates, guided by crystal symmetry principles and augmented with certain measured crystal tensors. From NDA, we expect the ``conventional" crystal tensors to be given by:
\begin{align}
 c^D_{nk} &\equiv \widetilde{c}^D_{nk} \frac{N_c \alpha}{a_0 V_c}, \label{eq:cNDA}\\
 h_{nk} &\equiv \widetilde{h}_{nk} \frac{N_c e a_0}{V_c}, \label{eq:eNDA} \\
   \beta_{nk}^S &\equiv \widetilde{\beta}^S_{nk} \label{eq:betaNDA}
\end{align}
with $N_c$ the number of atoms per unit cell. This NDA approach, which we have been unable to find in the literature, typically agrees with the measured (and/or calculated) values of the corresponding stiffness, piezoelectricity, and dielectric properties within an order of magnitude. The disagreement between the universal NDA approach and individual material properties is captured by the tensors $\widetilde{c}^D_{nk}$, $\widetilde{h}_{nk}$, and $\widetilde{\beta}^S_{nk}$, which generally have $\mathcal{O}(1)$ dimensionless coefficients. 


Crystal symmetry dictates that certain components of crystal tensors vanish. For example, $\widetilde{h}_{k n}$ and $\widetilde{\xi}_{k n}$ are odd under a parity transformation of the crystal lattice, so they must vanish identically for parity-even crystals. The piezoaxionic effect can therefore only be present in crystals with piezoelectric point groups, which constitute 20 geometric crystal classes (out of a total of 32)~\cite{nye1985physical}. The point group of the crystal further constrains relations among the components of the piezoelectric and piezoaxionic tensors. 
For example, for the crystal class 32 which includes quartz, the symmetry of the piezoelectric tensor is given by:
\begin{align}
    h_{nk} = \begin{pmatrix}
h_{11} & -h_{11} & 0 & h_{14} & 0 & 0\\
0 & 0 & 0 & 0 & -h_{14} & -h_{11} \\
0 & 0 & 0 & 0 & 0 & 0
\end{pmatrix},
\end{align}
and the $\xi$ tensor has exactly the same symmetry structure.
The electroaxionic tensor $\widetilde{\zeta}_{nk}$ is even under a parity transformation of the crystal structure, and is generally nonvanishing for any crystal.

Based on the crystal symmetries and the parity properties of the different coefficients, we expect that:
\begin{align}
     \left(\widetilde{\xi}_{nk}\right)^\intercal \sim \widetilde{h}_{nk} \quad \text{and} \quad \widetilde{\zeta}_{nk} \sim \widetilde{\beta}^S_{nk}, \label{eq:NDArelation}
\end{align}
allowing a preliminary estimate of the piezoaxionic and electroaxionic tensors from \emph{measurements} of piezoelectric and dielectric properties. The relations in Eq.~\ref{eq:NDArelation} capture suppression (enhancement) effects in the crystal, leading to numerically small (large) $\widetilde{h}_{nk}$ and $\widetilde{\xi}_{nk}$. In the case of quartz, for example, the piezoelectric tensor components are $e_{11} = -4.36\times10^9  \,\text{V}/\text{m} $ and $e_{14} = 1.04\times10^9  \,\text{V}/\text{m} $ at room temperature \cite{yang}, corresponding to $\widetilde{h}_{11} = -0.0609$ and $\widetilde{h}_{14} = 0.0145$ (these piezoelectric constants are related as $h_{mi} = \beta^S_{nm} e_{n i}$~\cite[Ch.~IIIA]{berlincourt1964piezoelectric}), suggesting that the piezoaxionic and electroaxionic tensors in the quartz crystal are also suppressed. (Quartz is not a good candidate material, primarily due to the smallness of the $\ce{^{29}_{14}Si}$ Schiff moment.) However, there do exist many families of strongly piezoelectric crystals, with large $\widetilde{h}$ tensors and thus likely unsuppressed $\widetilde{\xi}$ tensors.

We do not expect Eq.~\ref{eq:NDArelation} to hold much better than up to an $\mathcal{O}(1)$ number. Firstly, piezoelectricity depends on electron charge redistribution (under strain) in the entire unit cell, whereas piezoaxionicity only depends on electronic wavefunction changes near spin-polarized nuclei of interest. Secondly, only electron wavefunction components with minimal angular momentum (and thus large nuclear overlap) contribute to the piezoaxionic effect, whereas there is no such restriction for piezoelectricity.

In summary, we have shown that a nuclear Schiff moment can lead to stresses and electric fields via Eqs.~\ref{eq:Tstress} and~\ref{eq:Eelecfield}, respectively. QCD axion dark matter generates \emph{oscillatory} Schiff moments, so these stresses and electric fields also oscillate in time, and can in turn excite acoustic modes and currents, which we aim to detect with setups described in the next section. 

Finally, we provide a numerical estimate of the magnitude of the piezoaxionic effect. From Eq.~\ref{eq:Tstress}, the stress induced by a Schiff moment is equal to the stress induced by an equivalent strain of: 
\begin{align}
    &S_\text{eq} \sim \left|\left(\vect{c}^D\right)^{-1} \vect{\xi} \hat{\vect{I}} \overline{\theta}_a \right| \sim \frac{\widetilde{\xi} P}{\widetilde{c}^D } \frac{N_\mathsf{S}}{N_c} \frac{Z^2 \mathcal{R}}{(\nu_s \nu_p)^{3/2}} \frac{4\pi e}{\alpha a_0^3} \frac{\dd \mathsf{S}}{\dd \overline{\theta}} \overline{\theta}_a \label{eq:Sfid1} \\
    &\sim 3 \times 10^{-26} \, \frac{\widetilde{\xi} P}{\widetilde{c}^D}  \frac{N_\mathsf{S}}{N_c} \left(\frac{Z}{80}\right)^2 \frac{\mathcal{R}(\nu_s \nu_p)^{-\threehalf}}{10} \frac{\mathsf{S}}{\overline{\theta}_a \, e\,\mathrm{fm}^3}, \label{eq:Sfid2}
\end{align}
where we have assumed that the amplitude of $\overline{\theta}_a$ is that of Eq.~\ref{eq:thetaQCDaxion} and that $\mathsf{S}$ is linear in $\overline{\theta}_a$.
The fiducial equivalent strain of Eq.~\ref{eq:Sfid2} is just one order of magnitude smaller than the constraints achieved by the AURIGA collaboration, which set a limit on strains induced by scalar dark matter in an aluminum resonant-mass detector at the level of $S \sim \text{few} \times 10^{-25}$, based on a one-month data set~\cite{Branca:2016rez}.

\section{Experiment}\label{sec:experiment}
In Sec.~\ref{sec:theory}, we reviewed the mechanisms by which QCD axion DM can mix opposite-parity electronic states through short-range interactions with the nucleus, most notably via the nuclear Schiff moment induced by the irreducible QCD axion coupling. It is well known that the resulting effective Hamiltonian induces nuclear spin precession in parity-violating systems. This effect has been exploited in \emph{isolated} atoms or molecules, where parity is broken by a background electric field, to perform static measurements of the strong $\mathsf{CP}$ angle $\overline{\theta}$~\cite{mercuryEDMromalis}. Refs.~\cite{Graham:2013gfa, budker2014proposal} proposed using this nuclear precession effect combined with nuclear magnetic resonance techniques in solid-state systems, to look for an oscillatory $\overline{\theta}_a$ angle from QCD axion DM. Both of these aforementioned schemes rely on \emph{changes in the spin state}, specifically transverse polarization, as their primary signature.

The experimental setup we describe in this section utilizes the piezoaxionic effect of Eq.~\ref{eq:xidef} in piezoelectric crystals with spin-polarized nuclei. This setup does not rely on changes of the magnetization of the crystal. Instead, with the nuclear spin polarization fixed both in magnitude and direction, the oscillatory Schiff moment changes the electronic wavefuction so that the crystal lattice structure is no longer in equilibrium, resulting in a stress given by Eq.~\ref{eq:Tstress}. This axion-induced stress excites bulk acoustic modes in the piezoelectric crystal, whose electromechanical coupling admits an electric readout of this strain. Concurrently, the secondary electroaxionic effect parametrized by the $\zeta$ tensor of Eq.~\ref{eq:zetadef} is also generally present in any crystal, and produces an additional electric field contribution via Eq.~\ref{eq:Eelecfield}.  

Our proposed setup detects axion DM through the measurement of these bulk acoustic mode excitations. The acoustic modes are read out electrically (capacitively), and the resulting current is fed to a superconducting quantum interference device (SQUID). The main advantage of our proposal is the extraordinarily low \emph{mechanical} losses of certain piezoelectric crystals. This fact implies very high quality factors ($Q$-factors) of bulk acoustic modes, and low thermal noise. The associated mechanical $Q$-factors can be much larger than those of solid-state nuclear spin precession systems or those of purely electric modes. Furthermore, centimeter-scale crystals have natural mechanical resonance frequencies in the MHz range, providing natural amplification around all of the harmonics, with further scanning possible through electrical loading by an external circuit, as explained below. 

We sketch a simplified experimental setup in Sec.~\ref{sec:setup}, where we also discuss possible candidate materials. For concreteness, we quantitatively describe the signal for one particular geometry in Sec.~\ref{sec:signal}. We discuss different backgrounds and how they can be mitigated in Sec.~\ref{sec:backgrounds}. We finally present our sensitivity forecasts for our simplified setup with an idealized material in Sec.~\ref{sec:sensitivity}.

\begin{figure}[h!]
    \centering
    \includegraphics[width=0.48\textwidth, trim = 20 180 20 180, clip ]{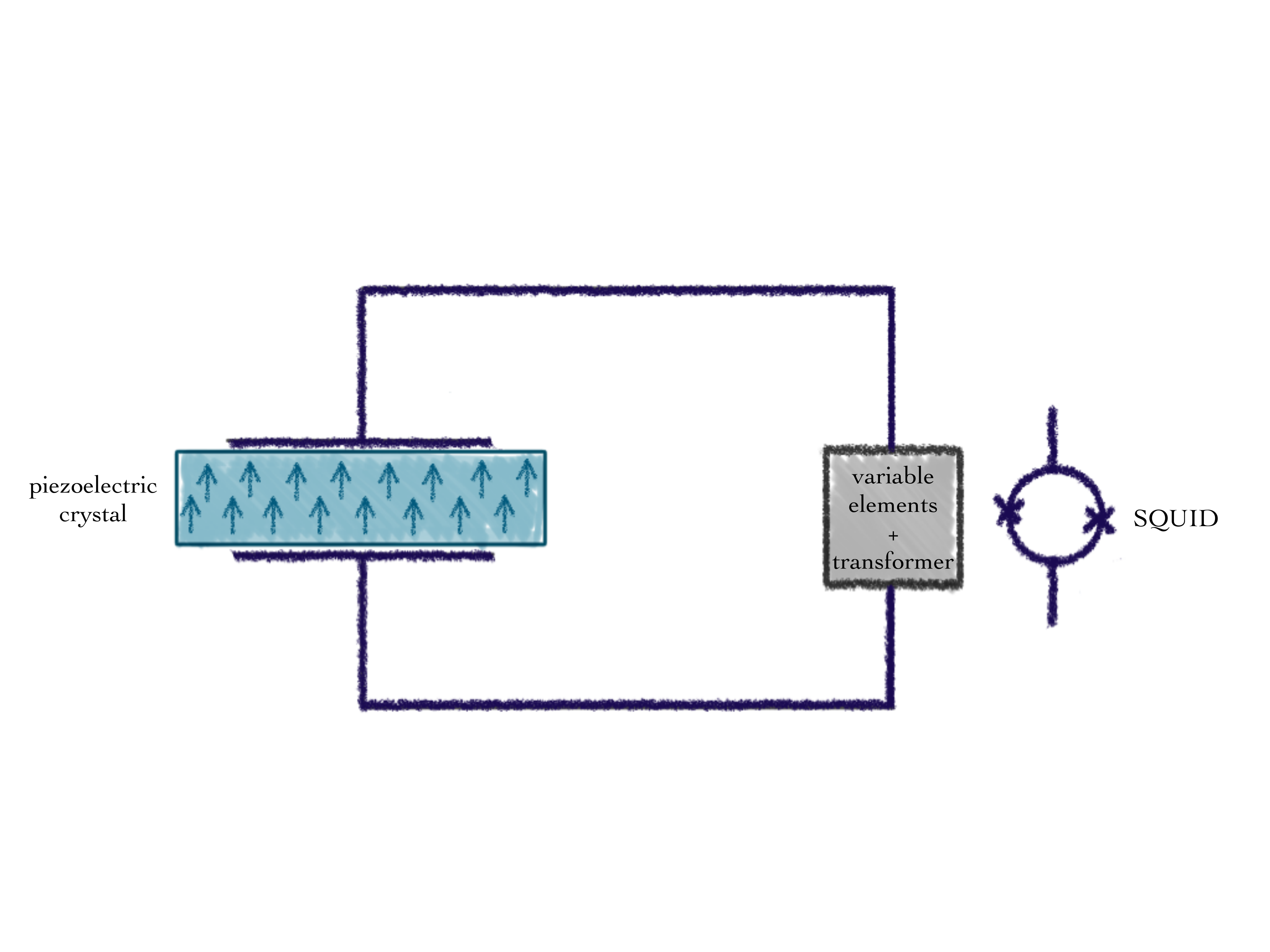}
    \caption{A simplified illustration of the proposed experimental setup. The axion-induced stress and resulting strain signal in a piezoelectric crystal with polarized nuclear spins is read out capacitively and ultimately measured through a SQUID. The circuit includes variable inductive and capacitive elements to scan around the mechanical resonance frequency, as well as a transformer circuit for impedance matching with the SQUID.}
    \label{fig:setup}
\end{figure}

\subsection{Setup}\label{sec:setup}
Our proposed setup consists of a piezoelectric crystal electroded on two of its faces, connected to a SQUID via a transformer circuit and variable input inductor and capacitor, as shown schematically in Fig.~\ref{fig:setup}. For simplicity of description, we will take the crystal to be a rectangular prism in the form of a thin plate, with the two opposing major faces electroded. The nuclear spins are aligned along an appropriately chosen crystal axis in order to maximize sensitivity. For nuclear spins, $\mathcal{O}(1)$ polarization can be achieved through the application of a large magnetic field at the cryogenic temperatures relevant for this experiment, since the Zeeman energy is of order the temperature $\mu_N B \approx 3.7\,\mathrm{mK}$ at $B = 10\,\mathrm{T}$. Due to the long relaxation times in solids, this polarization can be preserved for extended periods of time under application of a moderate magnetic field on the order of $10^{-4}$--$10^{-3}\, \mathrm{T}$~\cite{slichter_2011}.

We focus our treatment of the signal in Sec.~\ref{sec:signal} on the thickness expander mode, which has an acoustic wave vector and electric field perpendicular to the major faces; other geometries and modes are also possible and are discussed in App.~\ref{app:piezomodes}. Bulk acoustic modes in thin piezoelectric plates have been widely studied (see for example~\cite{Galliou2015, tichy2010fundamentals, zelenka1986}) and utilized in industrial, sensing, and fundamental physics applications~\cite{Goryachev2012,AURIGA-VIRGO-search}. We note that curvature of some of the crystal surfaces may help localize acoustic modes in the transverse direction and help minimize clamping losses~\cite{Goryachev:2012cq}, but for illustrative purposes we ignore this complication.

\begin{figure}[ht!]
   \begin{circuitikz}[american, /tikz/circuitikz/bipoles/length=1.1cm]
 \draw (0,2)  to[sV=$V_m$, voltage/american plus/.initial={}, voltage/american minus/.initial={}] (0,3.25) to[R=$R_m$]  (0,4.5)  to[L, l=$L_m$] (0,5.5) to[C=$C_m$] (0,6.5) -- (1.6,6.5)
 to[C=$C_c$] (1.6,2) -- (0,2);
  \draw (0.8,6.5) -- (0.8,7) -- (3.5,7) -- (3.5,6.5)
 to[twoport,t={$Z_\mathrm{BI}$}] (3.5,5) to[sV=$V_\mathrm{BI}$,voltage/american plus/.initial={}, voltage/american minus/.initial={}] (3.5,4) -- (3.5,2.5)  to[L, a=$L_1$] (3.5,1.3) to[C=$C_1$]
 (3.5,0.0) -- (0.8,0.0) to[sV=$V_{a}$, voltage/american plus/.initial={}, voltage/american minus/.initial={}] (0.8,2);
 \draw (4.2,2.7) -- (5.3,2.7) to[L, a=$L_i$] (5.3,1.2) -- (4.2,1.2) to[L, a=$L_2$] (4.2,2.7);
 \draw (6.2,2.7) to[squid, l=$\, L_\mathrm{SQ}$] (6.2,1.2);
 \draw [<->,>=stealth] (3.1,2.9)  to [bend left] node[pos=0.5,fill=white] {$M_{12}$} (4.6,2.9);
\draw [<->,>=stealth] (4.8,2.9)  to [bend left] node[pos=0.5,fill=white] {$M_i$} (6.4,2.9);
\draw[red,thick,dashed] (-1.25,1.75) -- (-1.25,6.75) node[fill=white] {$Z_\mathrm{crys}$} -- (2.75,6.75) -- (2.75,1.75) -- (-1.25,1.75);
\end{circuitikz}
    \caption{Equivalent electric circuit of the experimental set-up described in Sec.~\ref{sec:setup} and shown in Fig.~\ref{fig:setup}. The axion signal appears as an in-series voltage, $V_a$, to the piezoelectric crystal with capacitance $C_m$, inductance $L_m$, and resistance $R_m$ around mechanical resonance. Two of the crystal's faces are electroded, resulting in a clamped capacitance $C_c$. The electrodes are connected to variable electrical components $L_1$ and $C_1$. The current inside the circuit is read out inductively by a SQUID through a transformer. The SQUID and transformer contribute also a back-impedance $Z_\mathrm{BI}$ to the circuit (see text for more details).}
    \label{fig:equivalentcircuit}
\end{figure}
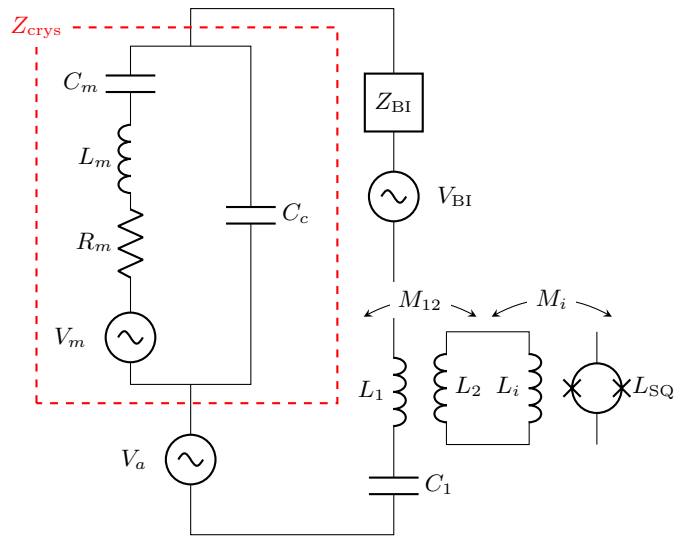

The axion signal and the electromechanical dynamics of the crystal will be further expounded in the next section, where we show that they can be modeled by an effective axion voltage $V_a$ in series with an effective ``primary'' impedance $Z_\mathrm{crys}$, respectively, as shown in the equivalent circuit of Fig.~\ref{fig:equivalentcircuit}. Near one of the acoustic resonances, the primary impedance can then further be approximated with an effective mechanical capacitance $C_m$, inductance $L_m$, and resistance $R_m$ (see App.~\ref{sec:piezoequiv}), all in parallel with the ``clamped capacitance'' $C_c$. The electrodes are connected to a variable capacitor $C_1$ and a variable inductor with self-inductance $L_1$. The latter has a mutual inductance $M_{12}$ to an inductor with self-inductance $L_2$, as part of a transformer circuit. The current inside the transformer circuit is sent to a SQUID with input inductance $L_i$, mutual inductance $M_i$, and SQUID self-inductance $L_\mathrm{SQ}$, as depicted in Fig.~\ref{fig:equivalentcircuit}. 
\begin{table}[!htbp]
    \centering
    \begin{tabular}{l l l}
    \hline \hline 
    Class & Candidates & Similar Crystals\\
    \hline
    \multirow{3}{*}{32}                     & $\mathrm{Na \vect{Dy}H_2 S_2 O_9}$        & $\mathrm{Si O_2}$ (quartz)\\
                                            &                                           & $\mathrm{Ga_5 La_3 Si O_{14}}$ (langasite)\\
                                            & $\mathrm{\vect{Bi}P O_4}$                 & $\mathrm{GaP O_4}$ (gallium orthophosphate) \vspace{2mm} \\  
    \multirow{2}{*}{$\mathrm{3m}$}    & $\mathrm{\vect{U} O F_4}$             & tourmaline\\
                                            & $\mathrm{\vect{U} Cd}$              & $\mathrm{Li Nb O_3}$ (lithium niobate) \vspace{2mm} \\
    \multirow{2}{*}{$\mathrm{4mm}$}         & $\mathrm{\vect{Dy} Si_3 Ir}$              & \multirow{2}{*}{$\mathrm{Li_2 B_4 O_7}$ (lithium tetraborate)} \\
                                            & $\mathrm{\vect{Dy} Ag Se_2}$              & \vspace{2mm} \\
    \multirow{2}{*}{$\mathrm{\bar{4}2m}$}   & $\mathrm{\vect{Dy} Ag Te_2}$              & $\mathrm{N H_6 P O_4}$ (ADP)\\
                                            & $\mathrm{\vect{Dy}_2 Be_2 Ge O_7}$        & $\mathrm{K H_2 P O_4}$ (KDP) \vspace{2mm}\\
                                            \multirow{1}{*}{$\mathrm{mm2}$}   & $\mathrm{\vect{U} CO_5}$                  & \multirow{1}{*}{$\mathrm{Ba_2NaNb_5O_{15}}$ (barium sodium niobate)}         \vspace{2mm}\\
    \hline \hline                                       
    \end{tabular}
    \caption{Candidate crystals for our setup, categorized by their crystal structure, as found in Ref.~\cite{Jain2013}. The nucleus with the largest Schiff moment in each compound is printed in bold. This table should be taken as an indicative but not exhaustive list.}
    \label{tab:materials}
\end{table}

\begin{center} \it{Materials} \end{center}
The sensitivity of our setup relies crucially on a number of material properties, most importantly piezoelectric and dielectric properties, low mechanical losses, and a high concentration of nuclei with large Schiff moments. In this section, we identify a number of suitable crystals with these properties, and discuss possible measures to improve the sensitivity of a given crystal. 

Before discussing candidate materials, we point out that since piezoelectric crystals are anisotropic, the orientation of different crystal cuts can result in significantly different material properties. These include variations in temperature coefficients, frequency stability, resonator quality factors, and spurious modes. For the example of quartz-type resonators, a comparison of state-of-the-art cuts and their properties can be found in Refs.~\cite{Salzenstein2016, goryachev:tel-00651960}, while Ref.~\cite{goryachev:tel-00651960} also presents a comparison of the properties of SC- and LD- cut resonators at cryogenic temperatures.  Non-contacting electrode designs where the electrodes are not deposited on the resonator itself can further reduce acoustic losses~\cite{Galliou2015,Goryachev2012, goryachev:tel-00651960}.

Piezoelectrics make up a large class of materials; of the 32 possible crystal symmetry classes, 20 exhibit piezoelectricity~\cite{IEEEstandard}. Nevertheless, the success of just a handful of piezoelectric crystals, for example quartz and langasite, has meant little previous need to explore the wide arena of potential materials. 
We provide in Tab.~\ref{tab:materials} an indicative subset of promising candidate materials collected from the database of The Materials Project~\cite{Jain2013}. The properties of these materials were predicted using density functional theory (DFT) and matched to an experimentally determined crystal structure. We searched for crystals with a large relative concentration of nuclei with known octupole-enhanced Schiff moments, and structural similarity to a well-developed piezoelectric material already used in bulk resonators (for examples of piezoelectric crystal classes and their properties, see Ref.~\cite{tichy2010fundamentals}). 

We expect that the ideal material will be comprised of a single crystal and not be ferromagnetic or strongly ferroelectric, so as to reduce losses associated with movement of domain walls. A number of optimal octupole-deformed nuclei were listed at the end of Sec.~\ref{sec:deformed}. In practice, the electronic properties of Eu (europium) and Gd (gadolinium) mean that their compounds are usually ferromagnetic. As a result, Tab.~\ref{tab:materials} mainly contains compounds of U (uranium) and Dy (dysprosium). We have also included crystals containing $\ce{^{209}_{83}Bi}$ (bismuth), as an example of a high-$Z$ nondeformed nucleus (Sec.~\ref{sec:nondeformed}). 

While DFT can be used to predict the unit cell volume, stiffness, dielectric, and piezoelectric properties of a candidate crystal with reasonable accuracy, the loss angles of the material (the imaginary parts of the crystal tensors) remain critical unknowns that must be determined experimentally. Nevertheless, one can speculate which materials could have low losses based on their similarity to known structures. Any resonator's quality factor can be further improved by increased purity, choice of cut, surface polishing, use of energy-trapping geometries, and lower temperature. It has been suggested that these measures can reliably improve the quality factor of many crystals by 3--4 orders of magnitude~\cite{Galliou2015,Goryachev2012}. While the exact choice of material for our setup is still to be determined, we think the requirements are not overly restrictive, and could be satisfied by a number of existing crystals.

\begin{table}[ht]
    \begin{tabular}{llll}
        \hline \hline
        Parameter & Value/Formula \qquad \qquad  & Parameter & Value/Formula \\ 
        \hline 
        $\rho$          & $12 \, \mathrm{g} \, \mathrm{cm}^{-3}$    & $V_c$             & $\mathrm{amu} \, A / \rho$ \\
        $N_\mathsf{S}$  & 5                                         & $[S_\Phi^\mathrm{SQ}]^{1/2}$ & $2.5 \times 10^{-7} \, \Phi_0 \, \mathrm{Hz}^{-1/2}$ \\
        $N_c$           & 10                                        & $L_\mathrm{SQ}$   & $0.1 \, \mathrm{nH}$ \\
        $Z$             & 92                                        & $R_\mathrm{SQ}$   & $10 \, \mathrm{\Omega}$ \\
        $A$             & 200                                       & $\eta_\mathrm{SQ}$ & $20$ \\
        $\mathsf{S}$    & $5 \,\overline{\theta}_a\,e\,\mathrm{fm}^3$ & $k_i, k_{12}$          & $0.75$ \\
        $n_N$           & $N_\mathsf{S}/V_c$                        & $L_i$             & $1 \, \mu\mathrm{H}$ \\
        $\mathcal{R}(\nu_s \nu_p)^{-\threehalf}$  & 10              & $C_1,L_1,L_2$             & variable \\
        $\mathcal{M}_1$ & $Z^2 \mathcal{R}(\nu_s \nu_p)^{-\threehalf} a_0^{-4}$         & $\delta_{L_1}$    & $10^{-6}$ \\
        $v^D$           & $\sqrt{\rho / c^D_{11}}$                  & $\delta_{C_1}$    & $0$ \\
        $c^D_{11}$      & $N_c \alpha / a_0 V_c$                    & $\delta_c$        & $10^{-9}$ \\
        $\beta^S_{11}$  & 1/3                                       & $\delta_\beta$    & $10^{-6}$ \\
        $h_{11}$        & $N_c e a_0 / V_c$                         & $\delta_h$        & $0$ \\
        $\zeta_{11}$    & ($4 \pi)^2 N_\mathsf{S} a_0^2 \mathcal{M} \mathsf{S} / V_c$   & $\hat{I}_1$ & $1$ \\
        $\xi_{11}$      & $4 \pi e N_\mathsf{S} \mathcal{M} \mathsf{S} / V_c $          & $T$   & $1\,\mathrm{mK}$ \\
 	\hline \hline
    \end{tabular}
    \caption{Fiducial parameters for an idealized setup assumed throughout this work. The definitions of each symbol can be found in the main text, primarily Sec.~\ref{sec:crystal} and Sec.~\ref{sec:signal}.}
    \label{tab:params}
\end{table}

\subsection{Signal}\label{sec:signal}
In this section, we describe the axion DM signal and how it excites the crystal's electromechanical modes, as well as the entire circuit. We focus on the thickness expander mode, with electric field parallel to the thickness direction (normal to the electrodes). It should be noted that the relevant indices for these modes will change depending on the crystal class; for concreteness, they have been chosen with quartz-type symmetry class 32 in mind.
Other possible modes and geometries are discussed in App.~\ref{app:piezomodes}, and could be useful for searching for axion signals at lower frequencies and with other crystal classes.

\subsubsection{Thickness expander mode of a thin plate}\label{sec:thicknessexpander}
Other than the excitation due to the piezoaxionic effect, the setup and treatment below are analogous to that of Ref.~\cite[Sec.~3.V.D]{berlincourt1964piezoelectric}. We will take the piezoelectric crystal to be a rectangular prism with side lengths $\ell_i$ of high aspect ratio (thin plate): $\ell_1 \ll \ell_2, \ell_3$. We are interested in the ``fast thickness expander mode'': an electro-acoustic wave with both the propagation and displacement in the thickness direction, i.e.~wavenumber $\vect{k} \propto \hat{\vect{x}}_1$ and displacement $\vect{u} \propto \hat{\vect{x}}_1$. Such a mode only has a normal strain $S_1 \neq 0$, while the other 5 strain components vanish identically. For a high-permittivity dielectric medium with negligible flux leakage, we have that the electric displacements $D_2$ and $D_3$ vanish on the minor faces, and because all gradients are in the $\vect{x}_1$ direction, they also vanish in the bulk. Furthermore, inside the insulating crystal, we have $\vect{\nabla} \cdot \vect{D} = 0$, so $D_1$ is spatially uniform. Since the major faces are electroded in our fiducial setup, we are interested only in the electric field in the thickness direction. The constitutive Eqs.~\ref{eq:TVoigt} and~\ref{eq:EVoigt} therefore reduce to:
\begin{alignat}{7}
T_1 &= + c^D_{11} && S_1 - h_{11} && D_1 - \xi_{11} && \hat{I}_1 \overline{\theta}_a; \label{eq:T1constitutive}\\
E_1 &= - h_{11} && S_1 + \beta^S_{11} && D_1 - \zeta_{11} && \hat{I}_1 \overline{\theta}_a. \label{eq:E1constitutive}
\end{alignat}

The equation of motion for this effectively one-dimensional acoustic mode is $\rho \ddot u_1 = \partial_1 T_{1} =c_{11}^D \partial_1^2 u_1$, the latter equality following from uniformity of $D_1$, $\hat{I}_1$, and $\overline{\theta}_a$. The solution to this equation that also satisfies the boundary conditions $T_1 = 0$ at $x_1 = 0, \ell_1$ is:
\begin{align}\label{eq:u1}
\hspace{-0.5em} u_1 &= \frac{h_{11} D_1 + \xi_{11} \hat{I}_1 \overline{\theta}_a}{\frac{\omega}{v^D} c^D_{11}} \left[\sin  \frac{\omega x_1}{v^D}  - \tan\frac{\omega \ell_1}{2v^D}  \cos\frac{\omega x_1}{v^D}  \right], 
\end{align} 
with $v^D = \sqrt{c^D_{11} / \rho}$ the crystal sound speed.
Above, the quantities $u_1$, $D_1$, and $\overline{\theta}_a$ are assumed to be oscillatory with angular frequency $\omega$; the factors of $e^{i \omega t}$ are not shown.

\begin{figure}
    \includegraphics[width=0.48\textwidth, trim = 10 0 0 0 ]{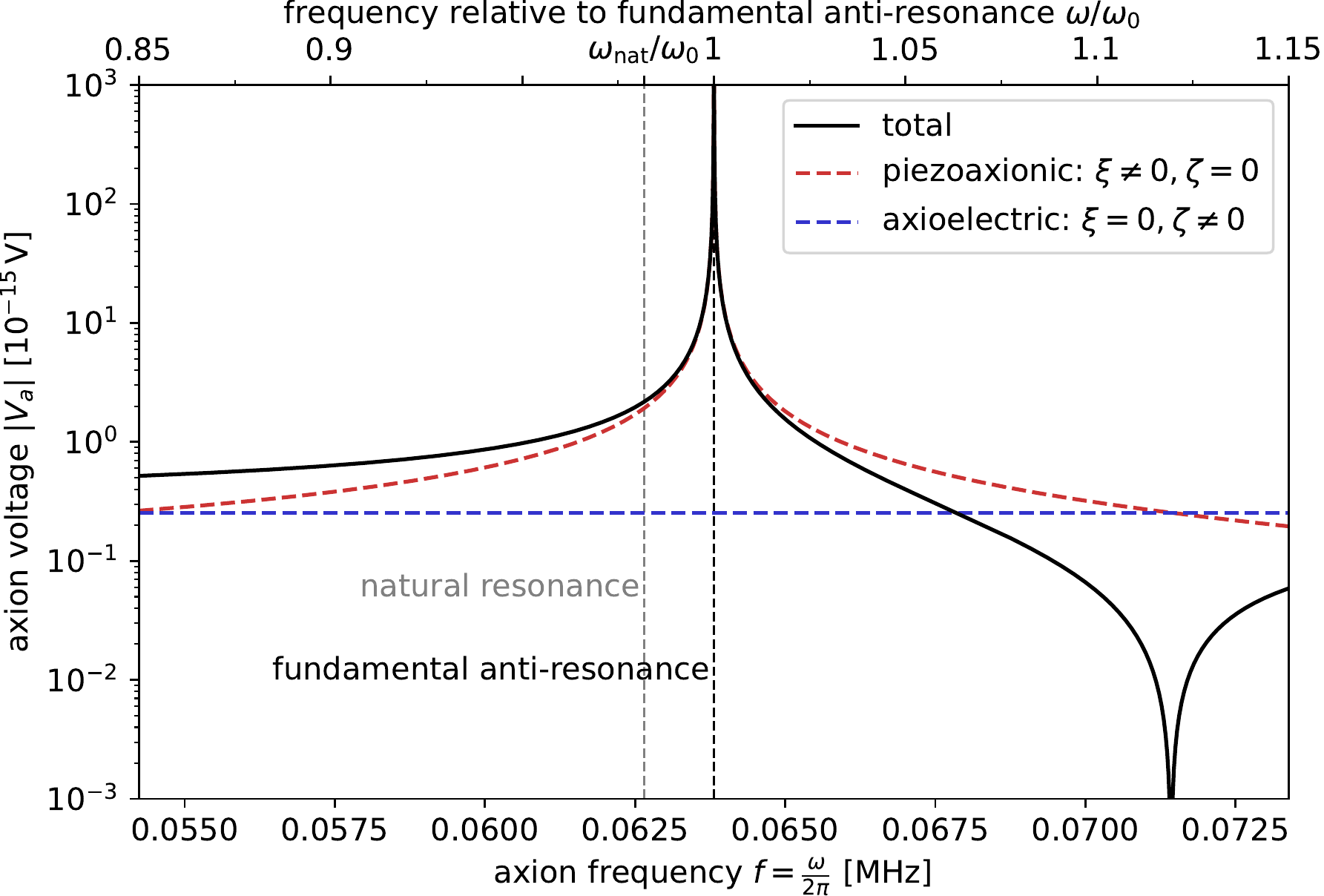}
    \includegraphics[width=0.48\textwidth]{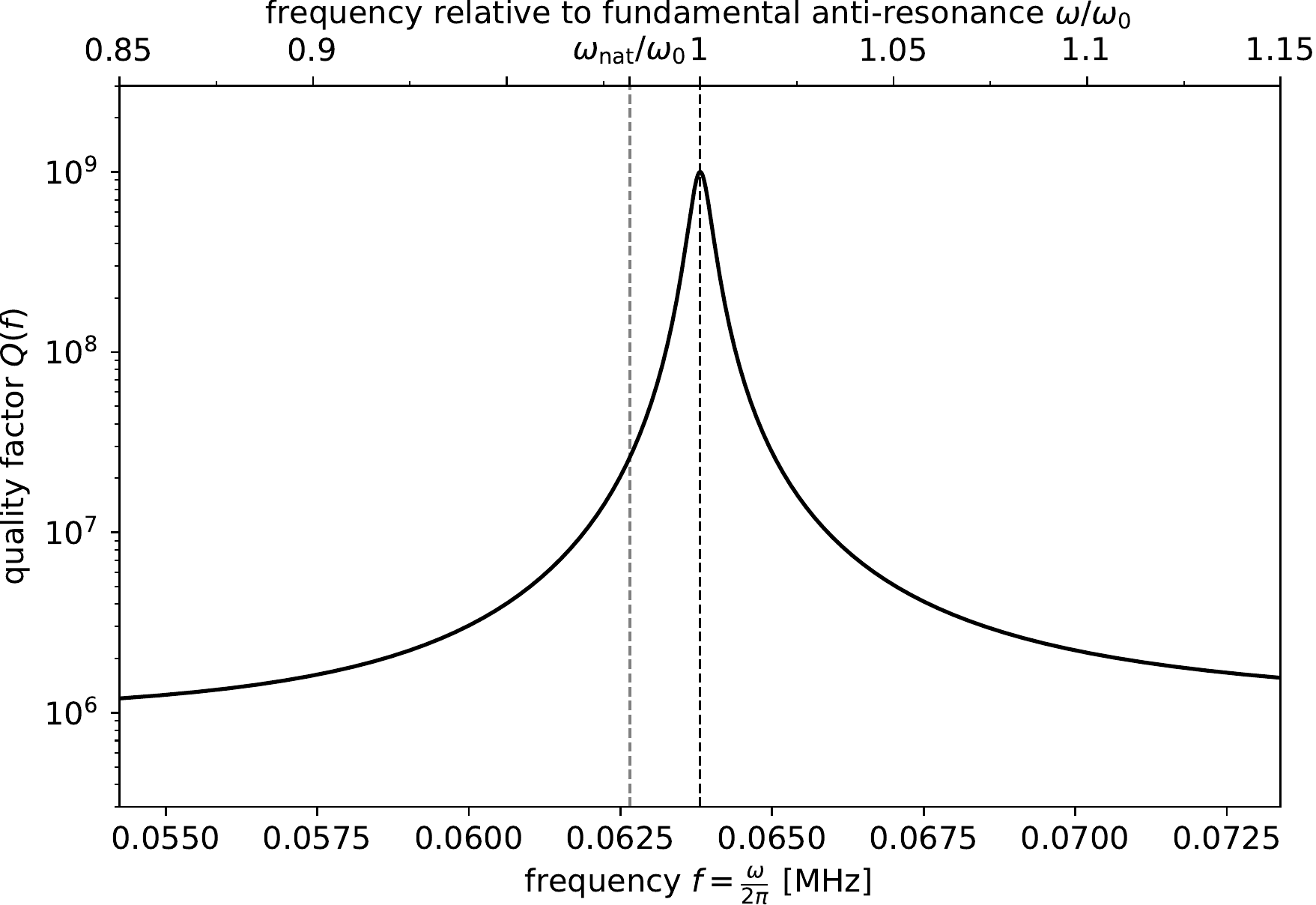}
    \caption{ {\it{Top panel:}} Axion induced voltage $V_a$ from Eq.~\ref{eq:Va} as function of frequency $f$ for the idealized crystal with parameters shown in Table~\ref{tab:params}. The piezoaxionic (blue dashed) and electroaxionic (red dashed) contributions are plotted separately, giving the total voltage in black. The piezoaxionic effect changes sign, and adds constructively (destructively) to the axioelectric polarizability effect just above (below) the fundamental mechanical resonance frequency (dashed black vertical line), for our choice of relative sign between the piezoaxionic and electroaxionic coefficients $\xi$ and $\zeta$. {\it{Bottom panel:}} Crystal quality factor $Q$ from Eq.~\ref{eq:Qfactor} as a function of frequency $f$ for the same idealized crystal. The fundamental and the natural resonance frequencies $f_0$ and $f_\mathrm{nat}$ are depicted by the black and dashed gray vertical lines, respectively. \nblink{supporting_plots_eta20.ipynb}}
    \label{fig:Va}
\end{figure}

The voltage difference
\begin{align}
    V = \int_0^{\ell_1} \dd x_1 \, E_1 = Z_\mathrm{crys} I + V_a \label{eq:Vtot}
\end{align}
measured across the electrodes attached to the crystal can be separated in a term equal to the current $I = i \omega \ell_2 \ell_3 D_1$ times the impedance
\begin{align}
    Z_\mathrm{crys} &= \frac{1}{i\omega C_c} \left[ 1 - k^2 \frac{2 v^D}{\omega \ell_1} \tan \frac{\omega \ell_1}{2 v^D}\right] \label{eq:Zcrystal}
\end{align}
and the axion-induced voltage 
\begin{align}
    V_a &= -\left\lbrace \frac{h_{11} \xi_{11}}{c_{11}^D} \frac{2 v^D}{\omega}  \tan \frac{\omega \ell_1}{2 v^D} + \zeta_{11} \ell_1 \right\rbrace \hat{I}_1 \overline{\theta}_a \label{eq:Va}
\end{align}
proportional to the nuclear spin polarization fraction $\hat{I}_1$ and axion-induced, oscillatory theta angle $\overline{\theta}_a$. In Eq.~\ref{eq:Zcrystal}, we have defined the clamped capacitance $C_c \equiv {\ell_2 \ell_3}/({\ell_1 \beta^S_{11}})$ and the electromechanical coupling factor:
\begin{align}
    k^2 \equiv \frac{h_{11}^2}{c^D_{11}\beta^S_{11}}. \label{eq:k2}
\end{align}
The axion-induced voltage $V_a$ thus appears \emph{in series} with that of the entire equivalent circuit of the crystal resonator. In the top panel of Fig.~\ref{fig:Va}, we plot in black the axion voltage amplitude, $V_a$, near the crystal's fundamental mechanical frequency $f \simeq f_0 \equiv v^D/ 2 \ell_1$ (vertical black dashed line). (The fundamental resonance frequency is sometimes also referred to as the ``anti-resonance frequency'', because it maximizes the effective impedance of the unloaded crystal.) We also define the natural resonance frequency $f_\mathrm{nat} = \omega_\mathrm{nat}/2\pi$ as the frequency at which the imaginary part of the crystal impedance $Z_\mathrm{crys}$ vanishes, and is depicted as the vertical gray dashed line in Fig.~\ref{fig:Va}. Below the fundamental frequency $f < f_0$, the $\xi$ and $\zeta$ contributions (plotted separately in red and blue, respectively) add constructively, while they add destructively above the mechanical resonant frequency. (This cancellation depends on the relative sign between $\xi$ and $\zeta$, which we have taken to be the same here; an opposite relative sign causes a cancellation below the fundamental resonance.) Near the fundamental resonance frequency, the voltage from the piezoaxionic effect dominates, due to the enhancement of the bulk acoustic mode displacement amplitude (Eq.~\ref{eq:u1}). This resonant enhancement turns out to be crucial to reach sensitivity to QCD axion DM parameter space.

Losses in the system are introduced through the complex stiffness and impermittivity tensors:
\begin{align}
    c^D_{11} &= \mathrm{Re}\left\lbrace c_{11}^D \right\rbrace ( 1 + i \delta_c), \\
    \beta^S_{11} &= \mathrm{Re}\left\lbrace \beta_{11}^S \right\rbrace ( 1 - i \delta_\beta);
\end{align}
where the loss angles $\delta_c$ and $\delta_\beta$ parametrize the mechanical and dielectric losses in the piezoelectric crystal. (For all quantitative results in this work, the signal response and noise treatment utilizes the above loss angles, not resistive elements such as $R_m$ from Sec.~\ref{sec:piezoequiv} and shown in Fig.~\ref{fig:equivalentcircuit}.) The effective quality factor $Q$ of the piezoelectric plate, as a circuit element referred to the primary circuit,  can be calculated as $Q(\omega) = (\text{stored energy})/(\text{energy dissipated per cycle})$. One then finds a frequency-dependent quality factor:
\begin{align}\label{eq:Qfactor}
    Q(\omega) = \frac{1}{\omega C_c \mathrm{Re}\left \lbrace Z_\mathrm{crys} \right\rbrace} \left[1 + k^2 \frac{1-\frac{3 v^D}{\omega \ell_1} \sin \frac{\omega \ell_1}{v^D}}{1+\cos \frac{\omega \ell_1}{v^D}} \right].
\end{align}
The quality factor is plotted in the bottom panel of Fig.~\ref{fig:Va}. We observe that, near the fundamental mechanical resonance frequency, the quality factor approaches the inverse of the  \emph{mechanical} loss angle $\delta_c$. Far away from this resonance, the quality factor asymptotes to the inverse dielectric loss angle $\delta_\beta^{-1}$, which is empirically much worse than the mechanical one for most materials. Thermal noise is therefore substantially reduced near the fundamental mechanical frequency of the system compared to that of a purely electrical resonator (Sec.~\ref{sec:backgrounds}).

\subsubsection{Readout circuit components}\label{sec:readout}
\begin{figure}
    \includegraphics[width=0.48\textwidth]{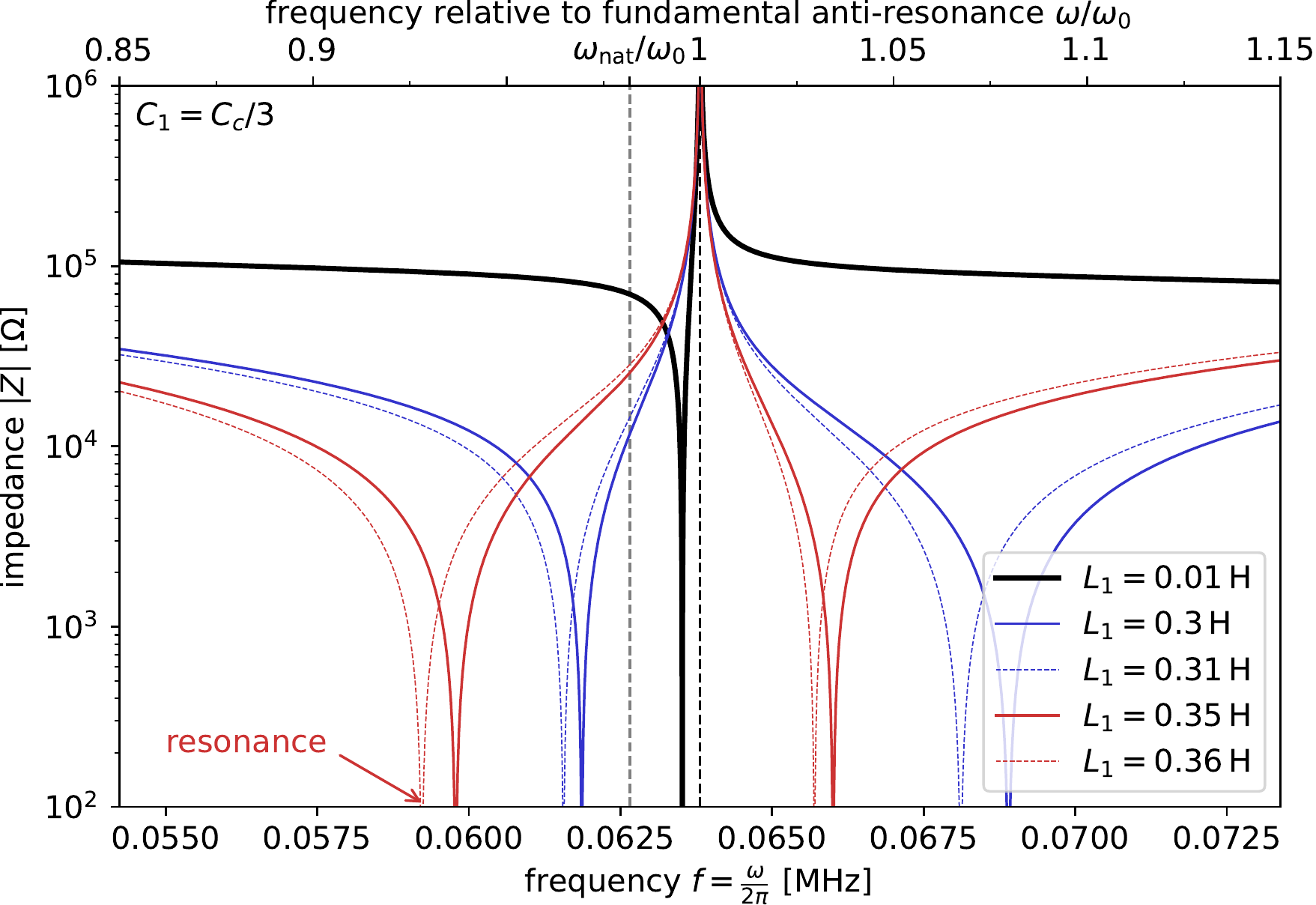}
    \includegraphics[width=0.48\textwidth]{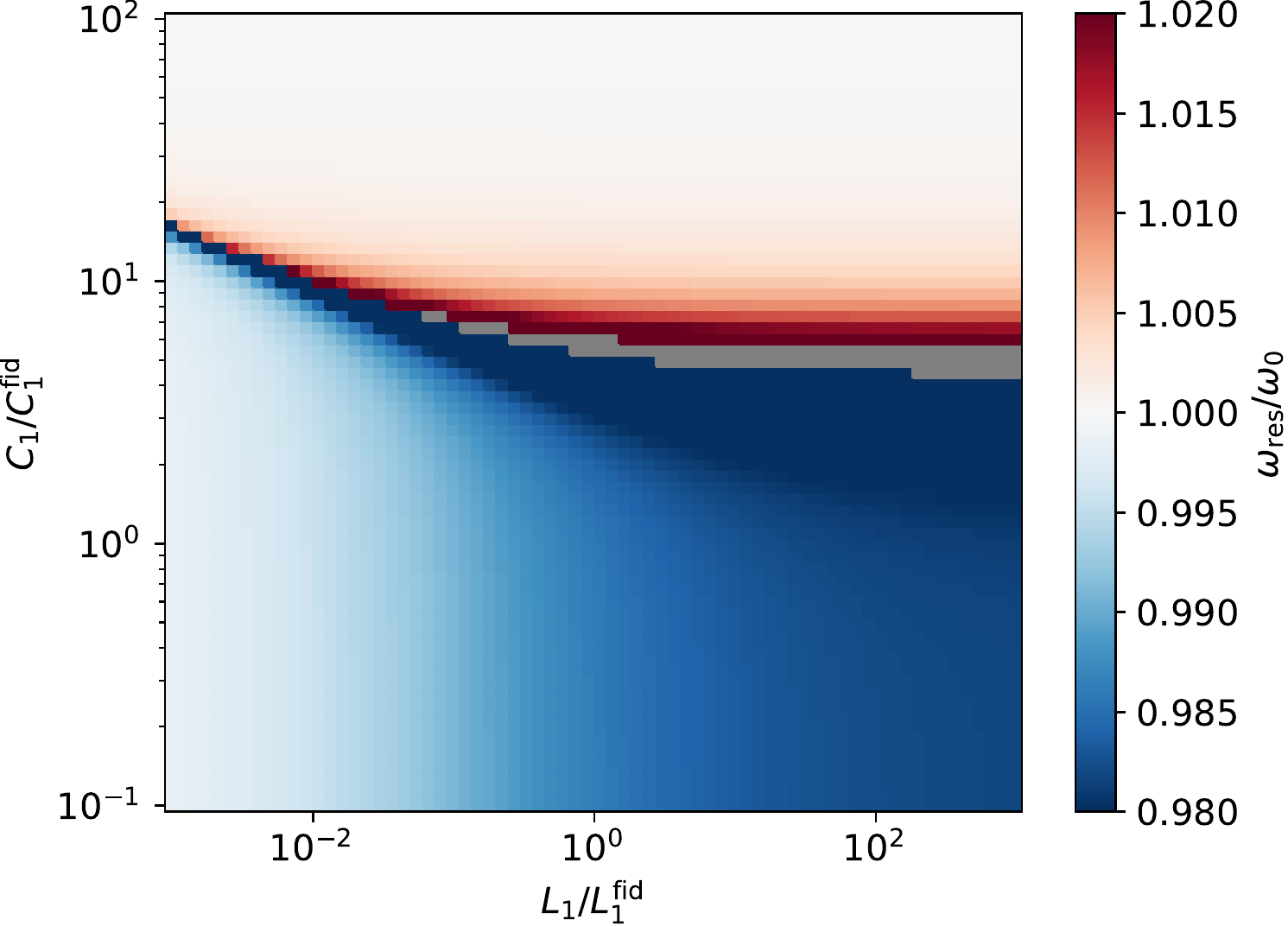}
    \caption{\textit{Top panel:} Total impedance as referred to the input circuit of Fig.~\ref{fig:equivalentcircuit}, for fixed $C_1 = C_c/3$ and different values of the transformer inductance $L_1$. For small $L_1 = 0.01\,\mathrm{mH}$ (black), the resonance frequency $f_\mathrm{res}$ below the fundamental mechanical resonance angular frequency $f_0$ (black dashed line) is shifted up above the natural resonance frequency $f_\mathrm{nat}$ (gray dashed line) due to the finite value of $C_1$. For increasing $L_1$, the resonance frequency is tuned downwards, as indicated by the blue lines for $L_1 = 12.0, 12.5\,\mathrm{mH}$. For yet larger inductances $L_1 = 16.0, 16.5\,\mathrm{mH}$, the resonance frequency associated with the mode \emph{above} the fundamental mechanical resonance approaches $f_0$ (red lines). \textit{Bottom panel:} Resonance angular frequency $\omega_\mathrm{res}$ (in units of $\omega_0$) as a function of $L_1$ and $C_1$. Blue (red) colors indicate the resonant branch below (above) $\omega_0$, like in the top panel. For both panels, the parameters of the idealized crystal used are given in Tab.~\ref{tab:params}. \nblink{supporting_plots_eta20.ipynb}}
    \label{fig:Z}
\end{figure}

Having described the signal and response of the piezoelectric crystal, we now turn to the other circuit components used for the readout of the signal due to the axion voltage of Eq.~\ref{eq:Va}, which are qualitatively similar to that of other experimental setups~\cite{chaudhuri2015radio, kahn2016broadband, ouellet2019design, goryachev2011, AURIGA-VIRGO-search}. The total impedance $Z_\mathrm{tot}$ of the circuit loop containing the piezoelectric crystal, which we shall refer to as the input circuit, is:
\begin{align}
    Z_\mathrm{tot} =  Z_\mathrm{crys} + \frac{1}{i \omega C_1} + i \omega L_1 + \Delta Z_\mathrm{BI} \label{eq:Ztot}
\end{align}
The second and third terms in $Z_\mathrm{tot}$ are the impedances of the capacitor with capacitance $C_1$, and inductor with self-inductance $L_1$, respectively. These two impedances serves as the main loads of the resonator, and can be used to control the setup's resonant frequency $f_\mathrm{res} = \omega_\mathrm{res} / 2\pi$, i.e.~the frequency at which $\mathrm{Im} \lbrace Z_\mathrm{tot} \rbrace$ vanishes, as shown in Fig.~\ref{fig:Z}. As discussed in Sec.~\ref{sec:sensitivity}, changing $L_1$ or $C_1$ is one of the methods that can be used to shift the resonance frequency of the circuit and scan (continuously) over possible axion masses. The transformer circuit and the SQUID contribute a fourth term to $Z_\mathrm{tot}$ in Eq.~\ref{eq:Ztot}, the back impedance~\cite{braginski_clarke_2006}:
\begin{align}
    \Delta Z_\mathrm{BI}  = \frac{ \omega^2 M_{12}^2 }{ i\omega L_2 + i \omega L_i + \omega^2 M_i^2 \left(\frac{1}{i\omega\mathcal{L}^r}+\frac{1}{\mathcal{R}^r}\right) },
\label{eq:ZBI}
\end{align}
where $\mathcal{L}^r$ and $\mathcal{R}^r$ are the reduced dynamical inductance and resistance of the SQUID, respectively~\cite{hilbert1985measurements}. This back impedance is typically negligible under optimal operating conditions.

The ultimate readout observable is the flux $\Phi_\mathrm{SQ}$ through the SQUID:
\begin{align}
    \Phi_\mathrm{SQ} = \frac{I \, M_{12} \, M_i}{L_i+L_2}, 
\end{align}
where $I$ is the input circuit current, which receives an additive contribution $I_a$ due to the axion voltage $V_a$ from Eq.~\ref{eq:Va}:
\begin{align}
    I_a = \frac{V_a}{Z_\mathrm{tot}} \label{eq:Ia}.
\end{align}

We take the mutual inductances to have fixed coupling coefficients $k_{12} \equiv M_{12} / \sqrt{L_1 L_2}$ and $k_i = M_i / \sqrt{L_i L_\mathrm{SQ}}$, respectively. While the flux through the SQUID (from the axion voltage) is maximized when $L_2 = L_i$, it is not generally optimal in view of back action noise (see Sec.~\ref{sec:squid}).  We note that the readout signal, for fixed axion voltage, is enhanced for two reasons in the above setup. Firstly, the axion signal current from Eq.~\ref{eq:Ia} is enhanced on resonance (when $|\mathrm{Im}\lbrace Z_\mathrm{tot}\rbrace|$ is small). Secondly, the transformer current steps up the current in the transformer circuit by a factor of $\mathcal{O}(\sqrt{L_1/L_2})$, resulting in a corresponding enhancement of the flux through the SQUID.

\subsection{Backgrounds}\label{sec:backgrounds}
In Sec.~\ref{sec:signal}, we computed how the axion DM background sourced the signal voltage $V_a$ in the input circuit, and how this propagated to a flux $\Phi_\mathrm{SQ}$ through the SQUID. In this section, we will outline the main backgrounds that compete with this signal. These noise sources can be divided into three categories: thermal noise due to dissipative circuit elements (Sec.~\ref{sec:thermal}), SQUID flux noise (Sec.~\ref{sec:squid}), and spin projection noise, i.e.~magnetization noise (Sec.~\ref{sec:spin}). For direct comparison to the axion voltage $V_a$,  we quantify all noise sources in terms of their equivalent voltage noise spectral densities:
\begin{align}
    S_{V}^\mathrm{tot} = S_V^\mathrm{th} + S_V^\mathrm{SQ} + S_V^\mathrm{spin} \label{eq:SVtot}
\end{align}
as referred to the input circuit, and assume a noise model with vanishing cross-correlations between voltage and current (see Ref.~\cite{chaudhuri2018optimal} for more discussion). We also comment on non-thermal vibration or seismic noise (Sec.~\ref{sec:seismic}), as well the limitations in cryogenic cooling arising from the possible use of metastable nuclei such as $\ce{^{235}_{92}U}$ and $\ce{^{237}_{93}Np}$ (Sec.~\ref{sec:heating}).

\subsubsection{Thermal Noise}\label{sec:thermal}

\begin{figure}
    \includegraphics[width=0.48\textwidth, trim = 20 0 0 0]{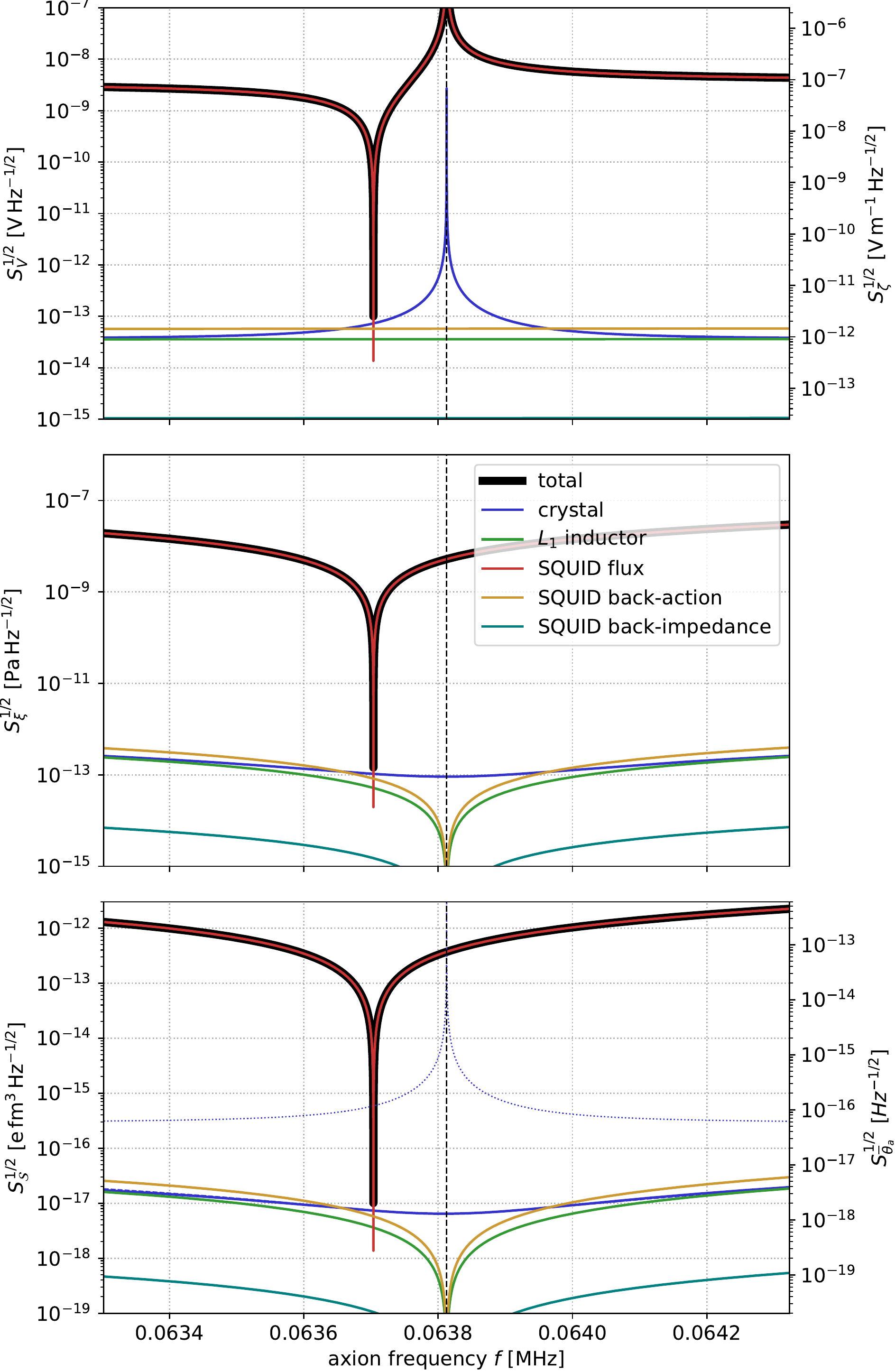}
    \caption{Amplitude spectral densities of all noise sources considered in the experimental setup with a crystal of dimensions $\ell_1 = 4\,\mathrm{cm}$ and $\ell_{2,3} = 40\,\mathrm{cm}$, variable capacitance $C_1 = C_c / 10$, and variable inductance $L_1 = 0.06\,\mathrm{H}$. All other parameters are the fiducial ones from Tab.~\ref{tab:params}. The top panel depicts noise sources expressed as equivalent voltage noise amplitudes $S_V^{1/2}$ of Eqs.~\ref{eq:SVtot},~\ref{eq:SVth}, and~\ref{eq:SVsquid}, 
    as referred to in the input circuit of Fig.~\ref{fig:equivalentcircuit}. Via Eq.~\ref{eq:Va}, they can be expressed as effective noise amplitudes for $\zeta_{11} \overline{\theta}_a$ (right axis of top panel) and $\xi_{11} \overline{\theta}_a$ (middle panel). In the bottom panel, they are shown as effective noise amplitudes for the Schiff moment $\mathsf{S}$ (left axis) and the QCD $\overline{\theta}_a$ angle (right axis), via Eqs.~\ref{eq:xidef} and~\ref{eq:zetadef}. \nblink{supporting_plots_eta20.ipynb}}
    \label{fig:noise}
\end{figure}

Thermal noise due to the lossy circuit elements is the primary source of noise on and near resonance. Its equivalent voltage noise spectral density $S_V^\mathrm{th}$ can be written as:
\begin{align}
    S_V^{\mathrm{th}} = 4 T \, \mathrm{Re}\left\lbrace Z_\mathrm{tot} \right\rbrace = S_V^{\mathrm{crys}} + S_V^{C_1} + S_V^{L_1} + S_V^{\mathrm{BI}}, \label{eq:SVth}
\end{align}
where $T$ is the thermodynamic temperature, and the breakdown in terms of different thermal noise sources is the same as in the sum of Eq.~\ref{eq:Ztot}. The thermal noise is proportional to the real part of the impedance, and thus to the loss angles $\delta_c$ and $\delta_\beta$ for the crystal circuit element, $\delta_{C_1}$ for the capacitor with complex capacitance $C_1 = |C_1|(1+i\delta_{C_1})$, and $\delta_{L_1}$ for the inductor with complex inductance $L_1 = |L_1|(1-i\delta_{L_1})$. The thermal voltage noise from the SQUID back impedance is linearly proportional to $\mathcal{R}^r$. In the top panel of Fig.~\ref{fig:noise}, we plot representative values for the thermal voltage noise $S_V^{\mathrm{crys}}$ of the crystal (blue), as well as $S_V^{L_1}$ of the inductor (green). The contribution $S_V^{\mathrm{BI}}$ of the back-action impedance of the SQUID readout system is negligible, and we assume the capacitance has a negligible loss angle (compared to that of $C_c$). Fiducial parameters for the setup are given in Tab.~\ref{tab:params}. The combined thermal noise is seen to dominate the total noise (black) on resonance. The total noise off-resonance is dominated by the SQUID noise to which we turn next.

\subsubsection{SQUID Noise}\label{sec:squid}

The voltage noise spectral density for the SQUID amplifier consists of current imprecision noise $S_{I}^\mathrm{flux}$ (due to the finite precision with which the flux can be read out) and back action noise $S_{V}^\mathrm{back}$:
\begin{equation}
S_{V, \text{tot}}^{\mathrm{SQ}} = |Z_{\text{tot}}|^2 S_{I}^\mathrm{flux} + S_{V}^\mathrm{back}.
\end{equation}
As demonstrated in Ref.~\cite[Apps.~E, F]{chaudhuri2018optimal}, under operating conditions with vanishing cross-correlated noise, these two noise sources are related by:
\begin{equation}
S_{V}^\mathrm{back} = \frac{\eta_\mathrm{SQ}^2 \omega^2}{S_{I}^\mathrm{flux}}, \label{eq:back}
\end{equation}
where $\eta_\mathrm{SQ} \geq 1$ determines the noise temperature $T^\mathrm{SQ}_N = \eta_\mathrm{SQ} \omega/\pi$ at each angular frequency $\omega$, and saturation of the inequality is achieved at the standard quantum limit (SQL). We take $\eta_\mathrm{SQ} = 20$ as a fiducial SQUID sensitivity, a value that has already been reached in a read-out circuit of a kHz-frequency mechanical resonator~\cite{falferi200810}.

The imprecision noise of the SQUID is related to the finite flux noise spectral density $S_\Phi^\mathrm{SQ}$, and quantifies how small a flux $\Phi_\mathrm{SQ}$ can be read out by the device. Translating this flux noise to an equivalent current noise in the input circuit gives:
\begin{align}
    S_{I}^\mathrm{flux} = \frac{(L_i+L_2)^2}{k_{12}^2 k_i^2 \,L_1 L_2 L_i L_\mathrm{SQ}} S_{\Phi}^\mathrm{SQ}. \label{eq:SVsquid}
\end{align}
As mentioned in Sec.~\ref{sec:signal}, the value of $L_2$ that maximizes the flux through the SQUID at fixed coupling coefficients $k_{12}$ and $k_i$ (and thus minimizes $S_{I}^\mathrm{flux}$) is $L_2 = L_i$. However, this flux maximization also maximizes the back action noise, cfr.~Eq.~\ref{eq:back}, and is typically suboptimal. Ref.~\cite{chaudhuri2018optimal}[App.~F] found the best operating point for any search for a near-monochromatic signal with log-uniform priors on its frequency; in the limit of $T \gg \omega$, which applies to our low-frequency setup, this corresponds to:
\begin{align}
S_{V}^\mathrm{back} = 2 T \, \mathrm{Re} \lbrace Z_\mathrm{tot} \rbrace, \label{eq:back2}
\end{align}
which we take to hold on resonance. For any resonance frequency of our system, we try to satisfy Eq.~\ref{eq:back2} by tuning the value of $L_2$. This could also be improved in future by using a readout that can perform back-action evading measurements, for example \cite{Kuenstner:2022gyc}.

The energy resolution of an optimized SQUID is limited by the uncertainty principle to satisfy the inequality~\cite{tesche1977dc}:
\begin{align}
    \frac{S_\Phi^\mathrm{SQ}}{2 L_\mathrm{SQ}} \gtrsim \pi,\label{eq:SPhiLim}
\end{align}
a limit that has been nearly saturated within a factor of $\mathcal{O}(1)$ by many groups~\cite{RevModPhys.92.021001}. For the purposes of our projections, we assume fiducial SQUID parameters from reference \cite{falferi200810} of $[S_\Phi^\mathrm{SQ}]^{1/2} =2.5 \times 10^{-7} \, \Phi_0 \, \mathrm{Hz}^{-1/2}$, $L_\mathrm{SQ} = 0.1 \, \mathrm{nH}$ and $L_i = 1 \, \mu\mathrm{H}$, where $\Phi_0 = \pi/e$ is the magnetic flux quantum. The SQUID's energy resolution is however not the sensitivity-limiting factor, which is instead the balancing of back impedance and back action and imprecision noise of Eqs. ~\ref{eq:ZBI},~\ref{eq:back} and ~\ref{eq:SVsquid}.

In the top panel of Fig.~\ref{fig:noise}, we plot the equivalent imprecision voltage noise of the SQUID flux noise in red (the first term in Eq.~\ref{eq:SVsquid}); off-resonance, imprecision noise is the sensitivity-limiting factor. In gold, we plot the corresponding back action voltage noise at the optimal operating point of Eq.~\ref{eq:back2}, which is always of order the thermal noise from the other circuit components.

\subsubsection{Spin Noise}\label{sec:spin}

For a crystal to exhibit the piezoaxionic effect due to the irreducible coupling of QCD axion DM, it needs to be populated by a high density of nuclear spins, whose fluctuations---spin projection noise or magnetization noise---can source an effective noise voltage in the circuit. The effective magnetic field noise spectral density of these spin fluctuations is~\cite{budker2014proposal}:
\begin{align}
    S_B^\mathrm{spin} = \frac{\mu_N^2 n_N}{8 \ell_1 \ell_2 \ell_3} \frac{T_2}{1+T_2^2 (\omega - \mu_N B_0)^2}, \label{eq:spinBnoise}
\end{align}
where $\mu_{N}$, $n_N$, and $T_2$ are the nuclear magnetic moment, the nuclear spin density, and the transverse spin relaxation time, respectively. There is generically a static magnetic field $B_0$ present in the material, from an externally applied field and/or from the aligned nuclear spins (unless they cancel each other). 
The resulting voltage noise spectral density into the input circuit can then be estimated as:
\begin{align}
    S_V^\mathrm{spin} = \omega^2 \ell_1^2 C_c^2 S_B^\mathrm{spin} \simeq \frac{\ell_2 \ell_3}{8 \ell_1 \left(\beta_{11}^S\right)^2} \frac{\mu_N^2 n_N}{T_2},\label{eq:SVspin}
\end{align}
where we have assumed $\omega \gg \mu_N B_0$.
The voltage amplitude spectral density of these spin fluctuations evaluates to $[S_V^\mathrm{spin}]^{1/2} \approx 1.0 \times 10^{-16} \, \mathrm{V} \, \mathrm{Hz}^{-1/2}$ for a magnetic moment equal to a nuclear magneton $\mu_N \equiv e / 2 m_p$, a typical solid-state transverse spin relaxation time $T_2 = 10^{-3} \, \mathrm{s}$, and otherwise the same parameters as used in Fig.~\ref{fig:noise}; this value is far below the other noise sources depicted in the top panel.

Magnetization noise can also arise from fluctuating magnetic impurities in the crystal. The electron impurity spin noise is similar to the one shown above, with $\mu_N$ substituted by the effective electron magnetic moment $\mu_e \sim \mu_B \equiv e/2m_e$ and correspondingly shorter transverse relaxation times $T_2'\sim 10^{-3}\, T_2$, since the electrons interact $\mu_e/\mu_N\sim 10^3$ times more strongly with their environment. Magnetic impurity concentrations as low as 1 part per billion have previously been reported \cite{Khatiwada_2015}. We find this source of noise to be subdominant to nuclear magnetization noise as long as magnetic impurities are kept to below 1 part per million, and certainly below the other noise sources plotted in Fig.~\ref{fig:noise}, even for much higher concentrations. Magnetization noise would however become the dominant source of broadband noise for any version of our proposed experiment with order-unity polarization fraction of \emph{electron} spins. For such magnetically ordered materials, magnetomechanical effects such as magnetostriction~\cite{Smith_1930} and piezomagnetism~\cite[Ch.~VII]{berlincourt1964piezoelectric}---lattice strain/stress proportional to magnetic fields---must also be considered in the noise budget, but are highly suppressed for the nonmagnetic crystals (with only nuclear spin polarization) considered in this work.

\subsubsection{Vibration Noise}\label{sec:seismic}

Non-thermal, external sources of vibration, such as those from seismic noise or human activity must be isolated from the crystal's acoustic modes. 
Around $f \sim \mathrm{kHz}$ frequencies, the displacement noise amplitude spectral density $S_\mathrm{disp}^{1/2}$ is empirically found to scale as~\cite{saulson1994fundamentals}:
\begin{align}
    S_\mathrm{disp}^{1/2} \sim \beta \, 10^{-13} \, \mathrm{m} \, \mathrm{Hz}^{-1/2} \, \left(\frac{\mathrm{kHz}}{f}\right)^2 
\end{align}
with $\beta$ a parameter ranging from $1$ to $10$ depending on the amount of surrounding human activity. In order to keep this vibration noise down to an equivalent strain noise amplitude of $S^{1/2}_\mathrm{strain} \sim 10^{-25} \, \mathrm{Hz}^{-1/2}$---as needed to detect the piezoaxionic effect from QCD axion DM (see Eq.~\ref{eq:Sfid2})---in a crystal of thickness $\ell \sim \mathrm{cm}$ at $f \sim \mathrm{MHz}$, one requires a vibration attenuation of at least $-160\,\mathrm{dB}$. In other words, the rms displacement of the crystal's major surfaces (off-resonance) needs to be suppressed by at least 8 orders of magnitude compared to the rms absolute displacement of the environment in the same frequency band. Such levels of vibration mitigation are well within the $-240\,\mathrm{dB}$ attenuation levels achieved by the AURIGA collaboration at $f \sim 900\,\mathrm{Hz}$~\cite{aurigasuspension}.

\subsubsection{Cooling Limitations}\label{sec:heating}
In order to achieve our target sensitivity, a crystal containing a high concentration of nuclei with octupole-enhanced Schiff moments needs to be cooled to cryogenic temperatures. Some candidate nuclear isotopes, such as $\ce{^{235}_{92}U}$ and $\ce{^{237}_{93}Np}$, are radioactive. The decay heat \cite{decayheattalk} from these metastable nuclei presents a challenge, as the cooling power of dilution refrigerators is limited. 

A dilution refrigerator's cooling power $\dot{Q}$ depends chiefly on the target temperature $T$ and the $\ce{^3He}$ flow rate $\dot{n}$~\cite{UHLIG1997279}:
\begin{eqnarray}
\dot{Q}\approx 8.4\,\mathrm{\mu W}~\left(\frac{\dot{n}}{10^{-3}\,\mathrm{mol/s}}\right)\left(\frac{T}{10\,\mathrm{mK}}\right)^2,
\end{eqnarray}
with a (high) fiducial flow rate indicated in brackets.

The largest crystal size $4\times 40\times 40~\text{cm}^3$ under consideration, with a density of $12\,\mathrm{g/cm^3}$, contains $64\,\mathrm{kg}$ of heavy (and potentially radioactive) nuclei, if they make up $10\,\mathrm{g/cm^3}$ in terms of mass density. The decay heating power of e.g.~$\ce{^{235}_{92}U}$ is approximately $60 \,\mu \mathrm{W/kg}$~\cite{decayheattalk}, which would preclude reaching temperatures below $200\,\mathrm{mK}$. This problem is even more severe for $\ce{^{237}_{93}Np}$, which has a decay heating power of $0.02\,\mathrm{W/kg}$. 

This decay heat makes the presence of metastable nuclei in a cryogenic environment challenging, if not prohibitive. As noted in Sec.~\ref{sec:nucleus}, there are several \emph{stable} nuclear isotopes with octupole-enhanced Schiff moments which do not suffer from this problem. Finally, it may also be possible to cool a single acoustic mode below the thermodynamic temperature of the crystal via optomechanical methods~\cite{o2010quantum}. Such futuristic experimental directions are outside the scope of this work.

\subsection{Sensitivity}\label{sec:sensitivity}
In Secs.~\ref{sec:signal} and \ref{sec:backgrounds}, we calculated the axion signal voltage $V_a$ (Eq.~\ref{eq:Va} and Fig.~\ref{fig:Va}) and the voltage noise spectral density $S_V^\mathrm{tot}$ (Eq.~\ref{eq:SVtot} and Fig.~\ref{fig:noise}), respectively, as referred to the input circuit. We will now synthesize these results to compute the sensitivity of our proposed experiment.

\begin{figure}
    \includegraphics[width=0.48\textwidth, trim = 10 0 0 0 ]{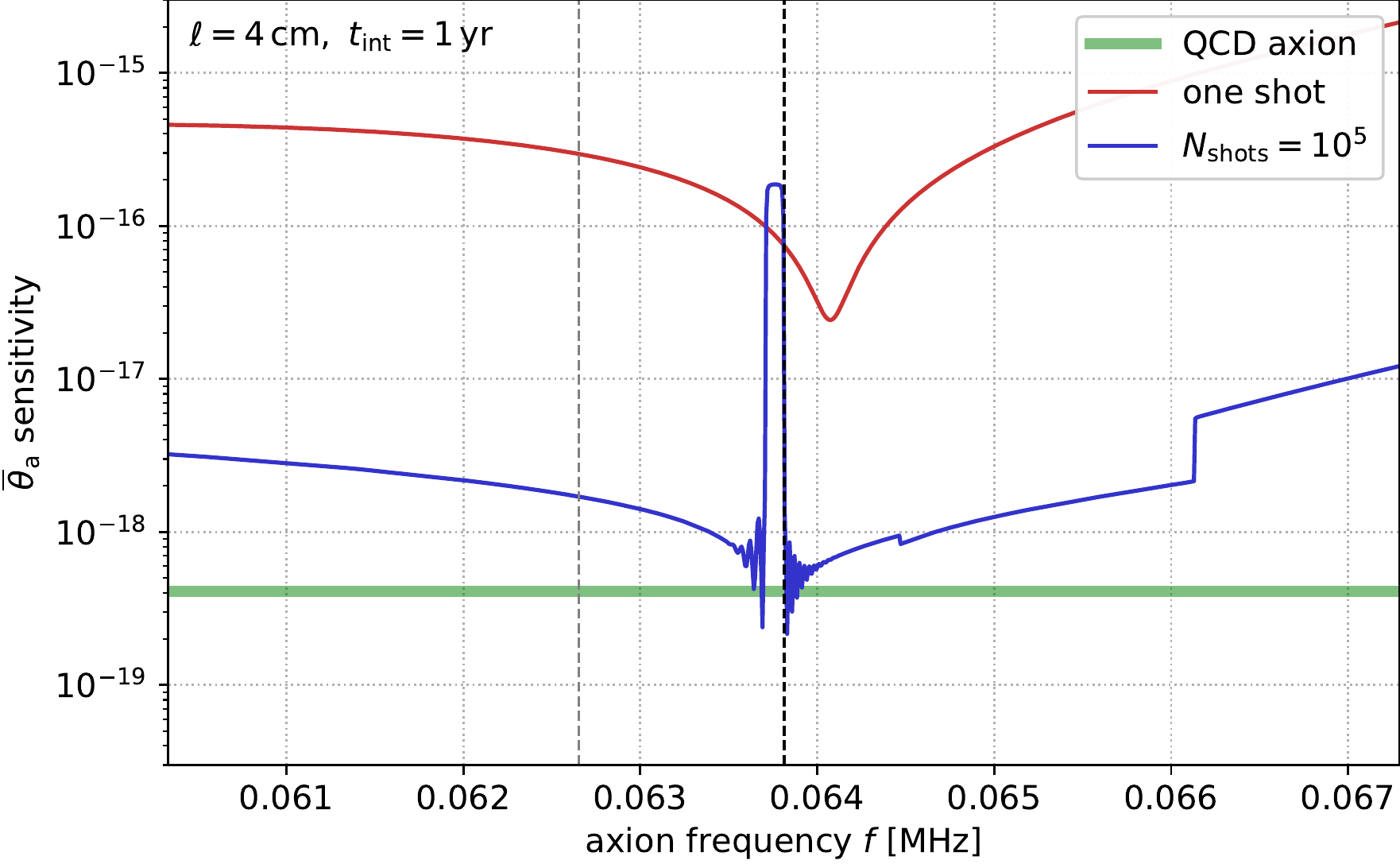}
    \caption{Sensitivity to an oscillatory theta angle $\bar{\theta}_a$ as a function of frequency for a single shot (red line) and for a scan around the mechanical resonance frequency of the first acoustic thickness expander mode of an idealized crystal with fiducial parameters given by Tab.~\ref{tab:params}. The green line shows the prediction for QCD axion DM. \nblink{sensitivity_eta20.ipynb}}
    \label{fig:SNRshot}
\end{figure}

After one ``shot'' of time $t_\mathrm{shot}$, which we assume to be larger than the coherence time $\tau_\mathrm{coh}$ of the axion DM signal in Eq.~\ref{eq:taucoh}, the signal-to-noise ratio (SNR) is:
\begin{align}
    \mathrm{SNR}_\mathrm{shot} &= \sqrt{\frac{|V_a|^2 \sqrt{t_\mathrm{shot} \tau_\mathrm{coh}}}{S_V^\mathrm{tot}}},
\end{align} 
where $S_V^\mathrm{tot}$ is to be evaluated at an angular frequency of $\omega = m_a$. 
The SNR of a single shot is largest whenever $|V_a|^2/S_V^\mathrm{tot}$ is largest, which typically only occurs for axion frequencies $f = m_a/2\pi$ in a small bandwidth around the resonance frequencies $f_\mathrm{res} = \omega_\mathrm{res}/2\pi$ of the complete circuit. This is shown in Fig.~\ref{fig:SNRshot}, which assumes a crystal of size $4\times 40\times 40~\text{cm}^3$ with parameters given in Table~\ref{tab:params}.  The red curve is a $\mathrm{SNR}_\mathrm{shot}=1$ contour expressed in terms of $\overline{\theta}_a$ sensitivity as a function of axion frequency. In addition to noise suppression, whenever a resonance is close to a fundamental resonance frequency of the crystal (the vertical black dashed line in Fig.~\ref{fig:SNRshot}), the sensitivity is enhanced further due to the increase of the axion voltage signal (see Fig.~\ref{fig:Va} and also the bottom panel of Fig.~\ref{fig:noise}).

\begin{figure*}
    \includegraphics[width=0.97\textwidth, trim = 10 0 0 0 ]{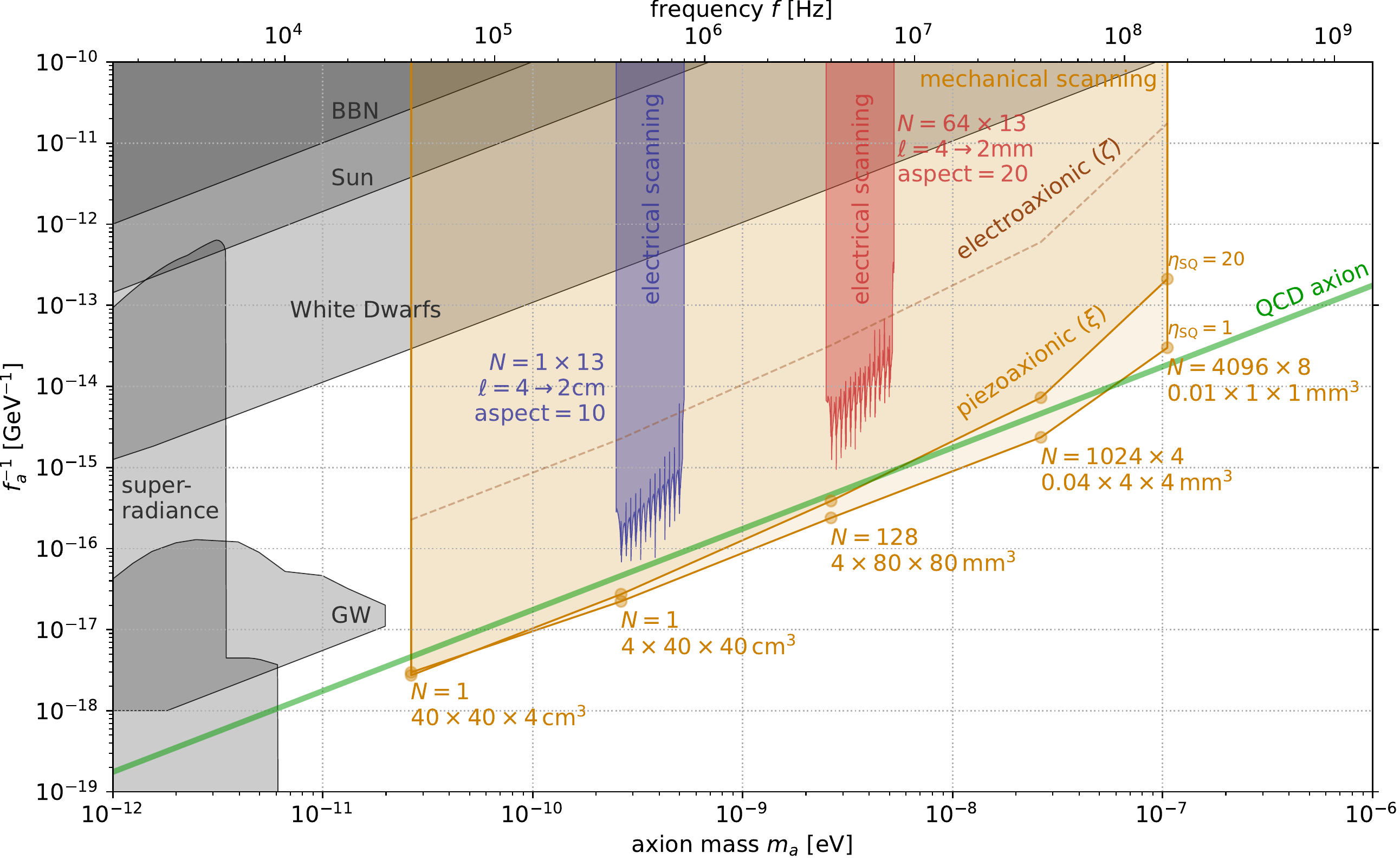}
    \caption{Axion parameter space probed by the setup described in Sec.~\ref{sec:experiment} and parameters given in Tab.~\ref{tab:params} as a function of the axion decay constant $f_a$ (inverse, vertical axis) and the axion mass $m_a$ (the top axis also indicates the corresponding frequency $f$). The blue and red shaded regions show the reach ($\mathrm{SNR_{tot}} = 1$ with Eq.~\ref{eq:SNRtot}) by scanning electrically around the first acoustic modes of a series of crystals with 13 different thicknesses. The gold region outlines the ultimate reach assuming mechanical scanning of the frequency for the parameters indicated. The ultimate reach, limited by a combination of SQUID noise and the crystal thermal noise, is shown for different values of $\eta_{SQ}$, the parameter indicating how close the SQUID operates to the quantum limit (see Sec.~\ref{sec:squid} and Sec.~\ref{sec:sensitivity} for more details).Similarly, the sensitivity achieved by the electroaxionic effect alone is outlined by the brown dashed line. The QCD axion relation between the mass and the decay constant (Eq.~\ref{eq:mafa}) is shown in green. The gray shaded regions are disfavored from BH superradiance~\cite{Baryakhtar:2020gao}, BBN~\cite{blum2014constraining}, the structure of the Sun~\cite{hook2018probing}, white dwarfs~\cite{Balkin:2022qer}, and neutron star binary mergers~\cite{zhang2021first}. \nblink{sensitivity_eta20.ipynb}}
    \label{fig:sensitivity}
\end{figure*}

The narrow instantaneous bandwidth of the setup necessitates a scanning strategy to achieve sensitivity to a broad range of axion masses. Our suggested strategy in this paper is to dial the resonance frequencies by changing the electric loads, namely the readout circuit's capacitance $C_1$ and inductance $L_1$ between each shot. For the resonance peak below the mechanical resonance, we take $C_1 = 10 \, |C_c|$, and $C_1 = 0.1 \, | C_c |$ for the resonance peak above the mechanical resonance frequency. We suggest scanning the resonance frequency with a large number $N = 10^4$ shots over regularly spaced steps in the range of $[2 \omega_\mathrm{nat} - \omega_0, 3 \omega_0 - 2 \omega_\mathrm{nat}]$ by changing the inductance $L_1$ in incremental steps. The value of the inductance needed lies in a range of a few orders of magnitude around the typical impedance-matched value of $|L_1| \sim 1/(\omega_0^2 |C_c| )$. The total SNR of all shots is computed as:
\begin{align}
    \mathrm{SNR}_\mathrm{tot} = \sum_\mathrm{shots} \left(\mathrm{SNR}_\mathrm{shot}^{-4}\right)^{-1/4}. \label{eq:SNRtot}
\end{align}
The isocontour $\mathrm{SNR}_\mathrm{tot}=1$ as a result of this procedure is plotted as the red line of Fig.~\ref{fig:SNRshot}. 

Figure~\ref{fig:sensitivity} shows in blue the forecasted sensitivity for 13 crystals with thicknesses $\ell_1$ logarithmically spaced between $4\,\mathrm{cm}$ and $2\,\mathrm{cm}$ and an aspect ratio of 10 (i.e.~transverse sizes $\ell_2 = \ell_3 = 40 \rightarrow 20\,\mathrm{cm}$), for a total integration time of $t_\mathrm{int} = 10\,\mathrm{yr}$, and other parameters matching those of Fig.~\ref{fig:SNRshot} and Tab.~\ref{tab:params}. Sensitivity to the QCD axion can be reached for such a setup over an octave in axion masses. The red region is a scaled-down version of the setup with 13 different values of $\ell_1 = 4 \rightarrow 2\,\mathrm{mm}$, 64 identical crystals hooked up in series, and an aspect ratio of $\ell_{2,3}/\ell_1 = 20$.

The gold region in Fig.~\ref{fig:sensitivity} is the parameter space that can plausibly be covered by setups similar to those previously described, assuming \emph{mechanical scanning} of frequency, which is not discussed in detail in this work. Mechanical scanning would allow taking full advantage of the high $Q$ factor around the resonance frequency. This plotted region is constructed by interpolating between the on-resonance sensitivity for 4 setups (indicated by gold circles) with dimensions $\ell_1 \times \ell_2 \times \ell_3$ equal to  $40 \times 40 \times 4 \, \mathrm{cm}^3$, $4 \times 40 \times 40 \, \mathrm{cm}^3$, $4 \times 40 \times 40 \, \mathrm{mm}^3$, and $0.01 \times 1 \times 1 \, \mathrm{mm}^3$, respectively. The third and fourth fiducial setups assume a large number of crystals in series and parallel at any one time, namely $64 \times 1$ and $512 \times 32 = 8192$. For each setup, we take $C_1 = 0.1 \, |C_c|$  and $L_1  = (\omega_0^2 | C_c| )^{-1}$. To cover an octave within a 10-year integration time, the shot time is assumed to be $t_\mathrm{shot} = t_\mathrm{int} / (Q_{\mathrm{res}} \ln 2)$ where $Q_\mathrm{res} \equiv \omega_\mathrm{res}/\Delta \omega$ is the inverse fractional bandwidth, and $\Delta \omega$ is the range of $\omega$ over which $S(\omega) < 4 S(\omega_\mathrm{res})$.
With those assumptions, sensitivity to the QCD axion may be attained over a mass range of $m_a \in \left[10^{-11}~\text{eV},~10^{-8}~\text{eV}\right]$ for a fiducial SQUID noise temperature determined by $\eta_\mathrm{SQ} = 20$, and over $\left[10^{-11}~\text{eV},~10^{-7}~\text{eV}\right]$ for a quantum-limited amplifier ($\eta_\mathrm{SQ} = 1$). Also shown is the reach due to the electroaxionic effect (dashed brown), which is weaker. The superiority of the piezoaxionic effect is due to the enhancement of the axion-induced voltage on resonance and the large $Q$ factor associated with mechanical oscillations of the crystal.

The green line in Fig.~\ref{fig:sensitivity} shows the QCD axion prediction, while gray regions indicate previously known constraints. Black-hole (BH) superradiance~\cite{Arvanitaki:2009fg,Arvanitaki:2010sy} is a process wherein the spontaneous/stimulated production of axions can extract a large amount of angular momentum from BHs; we take the updated constraints from Ref.~\cite{Baryakhtar:2020gao}, outlining parameter space that is inconsistent with spin measurements on five known BHs in x-ray binary systems. (These are lower bounds on $f_a$, as axion self-interactions can quench this process.) At sufficiently small values of $f_a$, the value of $\overline{\theta}_a$ could be so large that the neutron-proton mass difference at neutron freeze-out alters big-bang nucleosynthesis (BBN) predictions significantly~\cite{blum2014constraining}. The region labeled ``Sun'' is excluded by direct measurements of the Sun, as the in-medium reduction of the axion potential will generically destabilize the axion to  $|\overline{\theta}_a| \sim \pi$ at those small values of $f_a$~\cite{hook2018probing}. A similar destabilization can occur much more easily (for much larger $f_a$) in neutron stars, with the sourced axion field altering short-range forces within a double-neutron-star binary system---shown are corresponding gravitational-wave (GW) constraints from Ref.~\cite{zhang2021first}. Near nuclear saturation densities, as exhibited by supernova remnants and cores of neutron stars (e.g.~SN1987A and Cassiopeia A, respectively), axion production through the nuclear coupling is very efficient, leading to a bound of order $f_a \gtrsim 10^8\, \mathrm{GeV}$ (outside the plot range in Fig.~\ref{fig:sensitivity}), though the precise value of this bound is not yet settled~\cite[{\S}91]{Zyla:2020zbs}. As discussed in Sec.~\ref{sec:axion}, any bounds (and also our projections) above the QCD axion line prediction of Fig.~\ref{fig:sensitivity} are model-dependent due to the fine-tuning needed in this part of the parameter space. 

In summary, Fig.~\ref{fig:sensitivity} illustrates how the piezoaxionic effect can be used to search for the irreducible coupling of QCD axion DM. For other setups and proposals to search for this coupling, see Refs.~\cite{budker2014proposal,JacksonKimball:2017elr, abel2017search,chang2019axionlike, aybas2021search}. Because of derivative suppression, this is also a challenging mass range for detection concepts that search for other couplings of QCD axion DM~\cite{salemi2021search,gramolin2021search,zhang2021wisplc,ouellet2019first,JacksonKimball:2017elr}. The parameters of Tab.~\ref{tab:params} should be taken as indicative, with significant research and development needed to achieve the sensitivity curves of Fig.~\ref{fig:sensitivity} in practice. Nevertheless, we believe our idealized forecast demonstrates that the piezoaxionic effect can be a powerful probe of QCD axion DM.

\section{Other Axion Couplings} \label{sec:othercouplings}
In Secs.~\ref{sec:theory} and~\ref{sec:experiment}, we have focused on the irreducible coupling of QCD axion DM in Eq.~\ref{eq:axion-gluon-coupling}, and the resulting $\mathsf{P}$- and $\mathsf{T}$-violating effects. Generic axion-like particles (ALP), as well as the QCD axion~\cite{di2016qcd}, can have shift-symmetric couplings to fermions and photons:
\begin{align}
    \mathcal{L} \supset \frac{G_{a\gamma \gamma}}{4} a F_{\mu \nu} \widetilde{F}^{\mu \nu} - \sum_{f=p,n,e} \frac{G_{aff}}{2} \partial_\mu a \overline{\psi}_f \gamma^\mu \gamma^5 \psi_f, \label{eq:Lderivative}
\end{align}
with $\widetilde{F}^{\mu \nu} \equiv \epsilon^{\mu \nu \rho \sigma} F_{\rho\sigma} / 2$. The shift symmetry $[a \to a + \text{constant}]$ of the resulting action implies that all physical effects from these couplings are proportional to \emph{derivatives} of the axion field, i.e.~$\partial_\mu a$, where now $a$ can be a generic ALP or the QCD axion. We do not consider in detail the effects from the axion-photon coupling in this work, as our preliminary calculations indicate they are numerically small, and electrical resonator setups such as those of Refs.~\cite{chaudhurisnowmass2021,chaudhuri2015radio,kahn2016broadband,salemi2021search,ouellet2019design} are better suited.

For the axion-fermion couplings, one finds the following single-particle Hamiltonian:
\begin{align}
    H_{f} = \frac{G_{aff}}{2} \partial_\mu a \gamma^0 \gamma^\mu \gamma^5 \simeq -
    \frac{G_{aff}}{2} \vect{\sigma}_f \cdot \left[\vect{\nabla}a + \dot{a} \frac{\vect{p}_f}{m_f} \right], \label{eq:Hf}
\end{align}
with $\vect{\sigma}_f$, $\vect{p}_f$, and $m_f$ the spin, momentum, and mass of the fermion $f$, respectively. The nonrelativistic limit was taken in the second equality. 
The ``axion wind'' interaction, the $\mathsf{P}$-even and $\mathsf{T}$-odd operator proportional to $\vect{\nabla} a$ (and thus the axion velocity) in Eq.~\ref{eq:Hf}, has been proposed to search for ALPs and the QCD axion, by detecting changes in transverse magnetization from precessing nuclear~\cite{Graham:2013gfa} and electron spins~\cite{BARBIERI2017135}. 

In a parity violating medium, the $\mathsf{P}$-odd and $\mathsf{T}$-even operator proportional to $\dot{a}$ in Eq.~\ref{eq:Hf}, generically contributes to the internal energy density $U$ of Eq.~\ref{eq:int-energy} an additive correction:
\begin{align}
    U &\supset \sum_e^\text{unit cell} \frac{\langle H_e \rangle }{V_c} = -\frac{G_{aee} \dot{a}}{2V_c} \sum_e \Big\langle \psi_e \Big| \frac{\vect{p}_e \cdot \vect{\sigma}_e}{m_e} \Big| \psi_e \Big\rangle, \label{eq:int-energy-fermion}
\end{align}
since there is no symmetry that forbids such a nonzero contribution, in analogy to the piezoelectric effect. The axion-nucleon coupling $G_{aNN}$ yields numerically small effects in what follows, since nuclear contributions to the crystal lattice dynamics are suppressed by positive powers of $m_e/m_N$ and/or $R_0 / a_0$, so we focus on electronic contributions from hereon; the sum in Eq.~\ref{eq:int-energy-fermion} is over all electrons in the unit cell. Note that since the operator in Eq.~\ref{eq:int-energy-fermion} is $\mathsf{T}$-even, the electrons need \emph{not} be spin-polarized or unpaired, which is advantageous from the point of view spin-noise backgrounds (Sec.~\ref{sec:spin}). The internal energy density correction of Eq.~\ref{eq:int-energy-fermion} results in corresponding additive changes to the constitutive equations for the stress and electric field (e.g.~Eqs.~\ref{eq:TVoigt} and~\ref{eq:EVoigt}):
\begin{alignat}{5}
    T_n &= \frac{\partial U}{\partial S_n} &&\supset -\frac{G_{aee} \dot{a}}{2} \theta_n   && \equiv -\frac{G_{aee} \dot{a}}{2} \frac{\alpha N_c}{V_c} \widetilde{\theta}_n, \label{eq:Ttheta}\\
    E_n &= \frac{\partial U}{\partial D_n} &&\supset -\frac{G_{aee} \dot{a}}{2} \eta_n && \equiv -\frac{G_{aee} \dot{a}}{2} \frac{\alpha N_c}{V_c} \frac{e a_0^2}{\alpha} \widetilde{\eta}_n; \label{eq:Eeta}
\end{alignat}
where by NDA, the 2-tensors $\widetilde{\theta}_n$ and $\widetilde{\eta}_n$ should be of order unity. In Eqs.~\ref{eq:Ttheta} and~\ref{eq:Eeta}, we estimate that there are $N_c$ number of valence electrons, and that each matrix element in Eq.~\ref{eq:int-energy-fermion} is $\mathcal{O}(\alpha)$. Unlike in the case of the piezoaxionic and electroaxionic effects from a Schiff moment, where only $j=1/2$ electrons around a high-$Z$ nucleus contribute, here \emph{all} valence electrons participate.

\begin{figure}
    \includegraphics[width=0.48\textwidth, trim = 10 0 0 0 ]{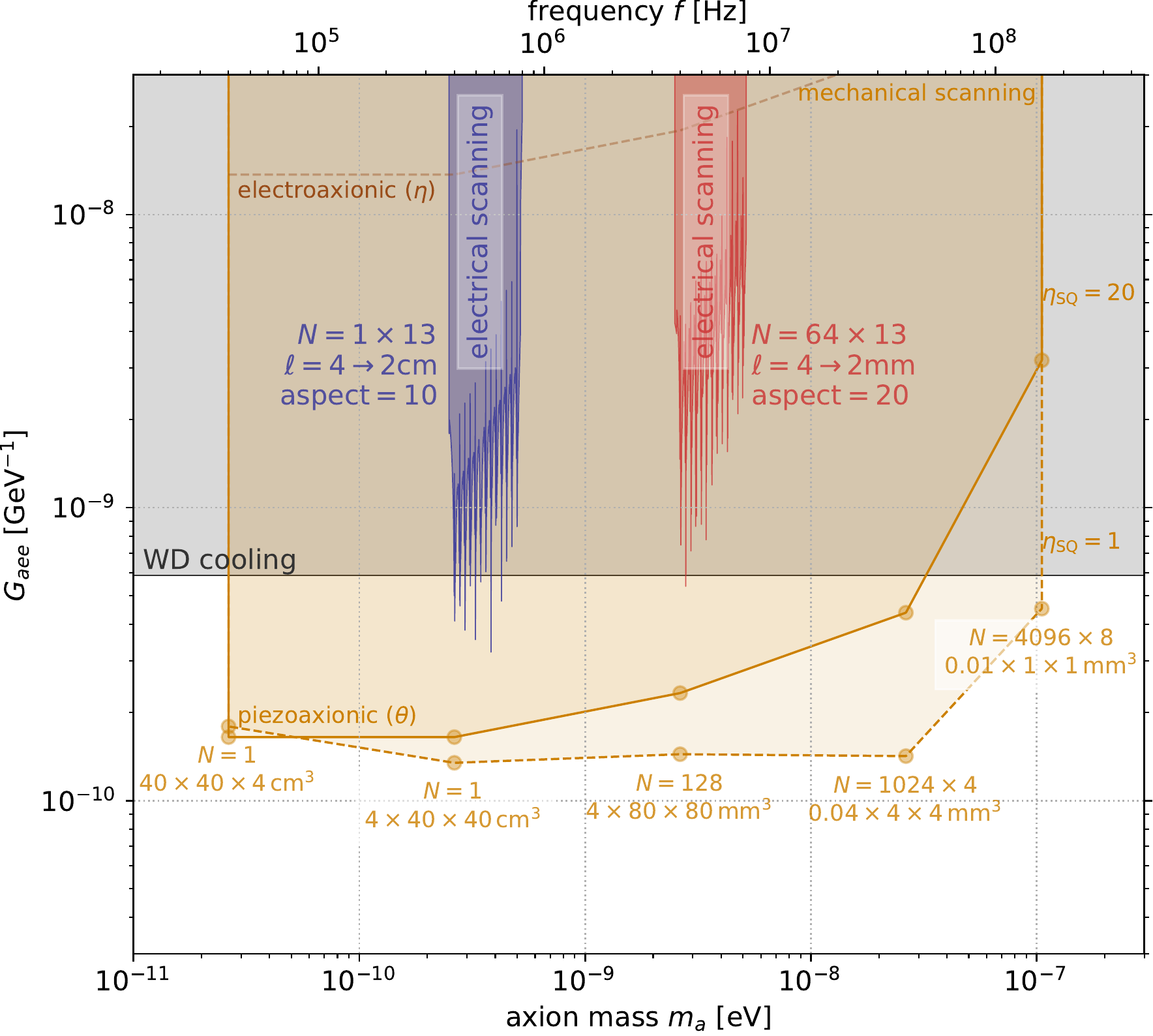}
    \caption{Sensitivity to the derivative axion-electron coupling $G_{aee}$ of Eq.~\ref{eq:Lderivative} as a function of mass $m_a$ of the axion-like particle. The setup and parameters are the same as in Fig.~\ref{fig:sensitivity}, with the additional assumption of $\widetilde{\theta}_1 = \widetilde{\eta}_1 = 1$ in Eqs.~\ref{eq:Ttheta} and~\ref{eq:Eeta}. The gray region indicates couplings strongly disfavored by cooling of white dwarf (WD) stars~\cite{bertolami2014revisiting}. \nblink{sensitivity_eta20.ipynb}}
    \label{fig:gaee-sensitivity}
\end{figure}

In Fig.~\ref{fig:gaee-sensitivity}, we show that with the same setup as described in the previous section, one is also sensitive to the axion-electron coupling of ALP DM. (The sensitivity calculation is completely analogous to the steps in Sec.~\ref{sec:experiment} with the appropriate substitutions of the axion-induced effects in Eqs.~\ref{eq:TVoigt} and~\ref{eq:EVoigt} with those of Eqs.~\ref{eq:Ttheta} and~\ref{eq:Eeta}, and is not repeated here.) The ultimate reach is up to one order of magnitude stronger than the bound $|G_{aee}| \lesssim 3 \times 10^{-13}/m_e$, obtained from considering the influence of axion-induced excess cooling on white-dwarf luminosity functions~\cite{bertolami2014revisiting}. While sensitivity to the irreducible coupling of the QCD axion requires a crystal hosting nuclei with large Schiff moments, a competitive reach to $G_{aee}$---and scalar DM couplings~\cite{arvanitaki2016sound}---can be already realized with commercially available crystal resonators such as quartz or gallium arsenide, and is thus an interesting target for a pilot experiment.

\section{Discussion}\label{sec:discussion}
In this work, we have described two new axion DM phenomena, and proposed an experimental concept that may discover QCD axion DM through its model-independent coupling over several decades in mass range. The setup relies on the intrinsic parity breaking in piezoelectric crystals. In such a material, an axion background results in a stress that can resonantly excite bulk acoustic modes if the axion frequency matches a fundamental acoustic frequency (or any of its harmonics) of the crystal, a phenomenon that we call the ``piezoaxionic effect". Due to the piezoelectric effect, this excited mode produces a voltage difference across the material that can be read out electrically. An axion background will also generically give another, non-resonant, additive contribution to this voltage difference---the ``electroaxionic effect". Our proposed setup is capable of detecting axions from tens of kHz to up to hundred(s) of MHz using well-known techniques from resonant-mass detectors developed for gravitational-wave detection, and low-noise electrical readout circuits used in other axion and dark-photon DM experiments. The most exciting signal is produced by the irreducible $\mathsf{P}$- and $\mathsf{T}$-violating coupling of the QCD axion to gluons, which furthermore requires that the piezoelectric crystal contain spin-polarized nuclei (to break $\mathsf{T}$ symmetry) with large Schiff moments. A detectable signal can also arise from the derivative coupling of axion-like particles to electrons, which is less challenging experimentally since the occurrence of the piezoaxionic effect for this coupling does not require polarized spins or the presence of special nuclear isotopes.

On the theoretical front, our treatment in Sec.~\ref{sec:signal} of the magnitude of the signal carries a large fractional uncertainty. The three bottlenecks are: a systematically improvable computation of nuclear Schiff moments proportional to $\overline{\theta}_a$ (Sec.~\ref{sec:nucleus}); an accurate determination of (soft) octupole deformation parameters for (meta)stable nuclear isotopes (Sec.~\ref{sec:deformed}); and a DFT calculation of the $\xi$ and $\zeta$ 3-tensors (Sec.~\ref{sec:crystal}), as well as the $\theta$ and $\eta$ 2-tensors for the ALP-electron coupling (Sec.~\ref{sec:othercouplings}). While difficult, these theoretical issues are not insurmountable, and we plan on addressing them in a future publication.

More research and development is also needed on the experimental front, regarding the following four issues in particular. Firstly, once a suitable crystal is identified using nuclear and crystal DFTs, its synthetic growth methods need to be studied and evaluated. Secondly, the elastic and electric loss angles $\delta_c$ and $\delta_\beta$ need to be experimentally determined at cryogenic temperatures, as well as optimal cut geometries and clamping methods to minimize external losses and optimize the signal. Thirdly, more investigation of frequency scanning strategies is desirable: this work showed that electric loading can effectively tune the resonance frequency (and retain sensitivity in a fractional bandwidth of order $k^2$ in Eq.~\ref{eq:k2}), but a (coarse) mechanical loading would reduce the total number of crystals needed to cover a fixed frequency range, and put less stringent requirements on the electrical components of the readout circuit ($L_1$, $C_1$, and the SQUID). Finally, it is worthwhile to study the feasibility of quantum-optics-based refrigeration of single acoustic modes (as opposed to the entire crystal), which would suppress the primary background (thermal noise) in our proposed setup and thus further enhance its sensitivity to this guaranteed signal of QCD axion DM. In a shorter time frame, and without these four advances, pathfinder experiments would already be sensitive to other dark matter candidates such as pseudoscalars coupled to electrons (Sec.~\ref{sec:othercouplings} and Fig.~\ref{fig:gaee-sensitivity}) and scalars coupled to electrons and/or photons~\cite{arvanitaki2016sound}.

This work studies the implementation of the piezoaxionic effect in the direct detection of DM; in forthcoming work, we will outline its potential applications to static searches for $\mathsf{P}$- and $\mathsf{T}$-violating phenomena. This new observable thus opens up exciting new avenues in the search for physics beyond the Standard Model.

\medskip

\acknowledgments{We thank Masha Baryakhtar, Sophie Beck, John Behr, Brian Busemeyer, Jonathan Engel, Peter Graham,  Matheus Hostert,  Junwu Huang, Andy Geraci, Maxim Goryachev, Kent Irwin, Andrew Jayich, Ania Jayich,  Dustin Lang,  Konrad Lehnert,  Dalila Pirvu, Lixin Sun, Giovanni Villadoro, and Neal Weiner for useful discussions. We are particularly grateful to Saptarshi Chaudhuri and Kent Irwin for insights into SQUID readout, and for pointing our omission of back-action amplifier noise in the first version of this manuscript. AA is grateful for the support of the Stavros Niarchos Foundation and the Gordon and Betty Moore Foundation. This material is based upon work supported by the National Science Foundation under Grant No.~2210551. The Center for Computational Astrophysics at the Flatiron Institute is supported by the Simons Foundation. Research at Perimeter Institute is supported in part by the Government of Canada through the Department of Innovation, Science and Economic Development Canada and by the Province of Ontario through the Ministry of Colleges and Universities.}

\bibliography{EDMbar}

\appendix

\section{Atomic Matrix Elements}\label{app:matrixel}
In this appendix, we elaborate on the calculation of the atomic matrix elements used in Sec.~\ref{sec:atom}. Symmetry dictates that to produce an expectation value for a $\mathsf{P}$-even observable such as an energy shift or mechanical stress, ones requires an even number of $\mathsf{P}$-violating perturbations. In other words, we know that a perturbation to an atomic orbital induced by a $\mathsf{P}$-violating potential from a nuclear Schiff moment, e.g.~$H_{\mathsf{S}}$, is not sufficient on its own to produce a $\mathsf{P}$-even effect at linear order by symmetry. The second $\mathsf{P}$-violating perturbation is sourced by the potential of the piezoelectric crystal lattice, $V_{\text{crys}}$. We assume $V_{\text{crys}}$ constitutes only a small correction to the atomic potential, such as in the tight-binding model for insulating materials~\cite{ashcroft2011solid}. 

For specificity, let us suppose that the ground-state wavefunction is $\ket{s^0}$, a mixed $s$-wave state with $j=1/2$ and equal admixtures of $m_j = \pm 1/2$, perturbed by the crystal potential as:
\begin{align}
    \ket{\widetilde{s}} &= \mathcal{C} \left( \ket{s^0}+\sum_{j, m_j} \ket{p_{j, m_j}^0}\frac{\bra{p_{j, m_j}^0}V_{\text{crys}}\ket{s^0}}{E_s^0-E_{p_{j, m_j}}^0}\right) \label{eq:epsilon1}\\
    &\equiv \epsilon_s \ket{s^0} + \sum_{j, m_j} \epsilon_{p_{j, m_j}} \ket{p_{j, m_j}^0}, \label{eq:epsilon2}
\end{align}
with $\mathcal{C} = 1-\sum_{j,m_j} \big|\bra{p_{j, m_j}^0}V_{\text{crys}}\ket{s^0}\big|^2 /( E_s^0-E_{p_{j, m_j}}^0)^2$. The superscript $^0$ denotes an unperturbed atomic wavefunction, and $\lbrace j, m_j \rbrace$ indicate the relativistic orbitals of the atomic $p$-level, i.e.~($j=1/2, m_j=1/2,-1/2$) and ($j=3/2, m_j=3/2,1/2,-1/2,-3/2$).
In practice, one can perform the above and following calculations for the $m_j = +1/2$ admixture of the $\ket{s^0}$ ground state, and then average the final result with the same for the $m_j = -1/2$ admixture.

The $\epsilon$ coefficients can be read off from matching Eqs.~\ref{eq:epsilon1} and~\ref{eq:epsilon2}. They can in principle be computed \emph{ab initio} within the framework of DFT, but care must be taken to compute atomic orbital projection coefficients in the presence of many valence electrons, and to ensure that the variation (with external strain or electric field/displacement) of these coefficients respects the point group symmetries of the crystal. This technically difficult calculation is left to future work, and instead we will use order-of-magnitude estimates based on experimentally measured quantities in Sec.~\ref{sec:crystal}.

We can now consider the leading-order energy shift when the atomic wavefunctions are perturbed by both the Schiff and crystal potentials as:
\begin{align}
    \braket{\widetilde{s} | H_{\mathsf{S}} | \widetilde{s}} = \sum_{{j, m_j}} \epsilon_s \epsilon_{p_{j, m_j}}^* \bra{s^0} H_{\mathsf{S}} \ket{p^0_{j, m_j}}+ \text{c.c.},
\end{align}
which directly leads to Eqs.~\ref{eq:<deltaH>} and~\ref{eq:renormalisedM} in the main text. The angular parts of the matrix elements between the spinor spherical harmonics $\Omega_{l,j,m_j}$ in Eq.~\ref{eq:renormalisedM} are:
\begin{alignat}{1}
    \bra{\Omega_{s,\half,+\half}}\hat{\vect{r}}\ket{\Omega_{p,\half,+\half}} &= -\frac{1}{3}\hat{\vect{z}}, \label{eq:ang1} \\
    \bra{\Omega_{s,\half,+\half}}\hat{\vect{r}}\ket{\Omega_{p,\half,-\half}} &= -\frac{1}{3}\hat{\vect{x}}+\frac{i}{3}\hat{\vect{y}}, \\ 
    \bra{\Omega_{s,\half,+\half}}\hat{\vect{r}}\ket{\Omega_{p,\threehalf,+\threehalf}} &=  -\frac{1}{\sqrt{6}}\hat{\vect{x}}-\frac{i}{\sqrt{6}}\hat{\vect{y}}, \\
    \bra{\Omega_{s,\half,+\half}}\hat{\vect{r}}\ket{\Omega_{p,\threehalf,+\half}} &= +\frac{\sqrt{2}}{3}\hat{\vect{z}}, \\
    \bra{\Omega_{s,\half,+\half}}\hat{\vect{r}}\ket{\Omega_{p,\threehalf,-\half}} &=  +\frac{1}{3\sqrt{2}}\hat{\vect{x}}-\frac{i}{3\sqrt{2}}\hat{\vect{y}}; \label{eq:ang5}
\end{alignat}
and similarly for the $l=0,m_j=-1/2$ admixture:
\begin{alignat}{1}
    \bra{\Omega_{s,\half,-\half}}\hat{\vect{r}}\ket{\Omega_{p,\half,+\half}} &= -\frac{1}{3}\hat{\vect{x}}-\frac{i}{3}\hat{\vect{y}}, \\
    \bra{\Omega_{s,\half,-\half}}\hat{\vect{r}}\ket{\Omega_{p,\half,-\half}} &= +\frac{1}{3}\hat{\vect{z}}, \\ 
    \bra{\Omega_{s,\half,-\half}}\hat{\vect{r}}\ket{\Omega_{p,\threehalf,+\half}} &= -\frac{1}{3\sqrt{2}}\hat{\vect{x}}-\frac{i}{3\sqrt{2}}\hat{\vect{y}}.  \\
    \bra{\Omega_{s,\half,-\half}}\hat{\vect{r}}\ket{\Omega_{p,\threehalf,-\half}} &= +\frac{\sqrt{2}}{3}\hat{\vect{z}}, \\
    \bra{\Omega_{s,\half,-\half}}\hat{\vect{r}}\ket{\Omega_{p,\threehalf,-\threehalf}} &= +\frac{1}{\sqrt{6}}\hat{\vect{x}}-\frac{i}{\sqrt{6}}\hat{\vect{y}},
\end{alignat}
Transitions with $|\Delta j| \geq 2$ or $|\Delta m_j| \geq 2$ are forbidden by selection rules.

\section{Long-wavelength Reduction}\label{app:long}

In this appendix, we derive Eq.~\ref{eq:int-energy} as the long-wavelength description of the crystal, reduced from the full energy functional that includes short-wavelength degrees of freedom, which are ``integrated out". We also derive how the piezoaxionic tensor $\xi$ and the axioelectric tensor $\zeta$ relate to wavefunction coefficients that can be computed within DFT and to the atomic matrix elements of Sec.~\ref{sec:atom}. Our treatment of the short-wavelength modes is based on Refs.~\cite{deGironcoli1989, Baroniphonons2001, tiersten1969linear}.

Denote the position of atom $I = ( l, s )$ by
\begin{align}
R_\alpha^{I} \equiv R_\alpha^l + \tau_\alpha^s + u_\alpha^{ls},
\end{align}
where $\vect{R}^l$ is the position of the $l$th unit cell in the Bravais lattice, $\vect{\tau}^s$ is the relative position of the $s$th atom within the unit cell, and $\vect{u}^{ls}$ is the out-of-equilibrium deviation. In what follows, vector indices will be $\lbrace \alpha, \beta, \gamma, \dots\rbrace $ subscripts running over the 3 spatial directions, unit cell labels $\lbrace l, m\rbrace$ superscripts running over $N$ unit cells with volume $V_c$, and atomic labels within the unit cell $\lbrace s, t \rbrace$ superscripts. 

We can express the total internal energy density around equilibrium to quadratic order in deviations, namely a homogeneous strain $S_{\alpha \beta}$ and electric displacement vector $D^\alpha$ as well as the individual atomic displacements $u_\alpha^{ls}$, as:
\begin{align}
&U = \label{eq:Ucomp}\\
&+\frac{1}{2} c^0_{\alpha \beta, \gamma \delta} S_{\alpha \beta} S_{\gamma \delta}  -  h^0_{\alpha,\gamma \delta} D_\alpha 
S_{\gamma \delta} + \frac{1}{2}  \beta_{\alpha \beta}^\infty D_\alpha D_\beta \nonumber \\
&+ \frac{1}{2 N V_c} \sum_{lm,st} C^{ls,mt}_{\alpha \beta} u_\alpha^{ls} u_\beta^{mt}\nonumber \\
&- \frac{e}{N V_c} \sum_{l,s} Z^{s}_{\alpha \beta} u^{ls}_\alpha D_\beta + \frac{1}{N V_c} \sum_{l,s} G^s_{\alpha,\gamma \delta} u^{ls}_\alpha S_{\gamma \delta} \nonumber \\
 &  -  \zeta^0_{\alpha \beta} \hat{I}_\alpha \overline{\theta}_a D_\beta - \sum_{s} \xi^{0,s}_{\alpha, \gamma \delta} \hat{I}_{\alpha}^s  \overline{\theta}_a S_{\gamma \delta} \nonumber \\
 &- \frac{e}{N V_c} \sum_{l,s} W_{\alpha \beta}^s u_\alpha^{l s} \hat{I}_\beta^s  \overline{\theta}_a. \nonumber
\end{align}
In the first two lines of the above equation, we have defined the bare elastic tensor $c^0_{\alpha \beta, \gamma \delta}$, the bare piezoelectric tensor $h^0_{\alpha, \gamma \delta}$, the bare dielectric impermittivity tensor $\beta^\infty_{\alpha \beta}$, the interatomic force matrix $C^{ls,mt}_{\alpha \beta}$, the effective charge matrix $Z^{\alpha \beta}_{s}$, the internal strain tensor $G^s_{\alpha, \gamma \delta}$, and homogeneous electric displacement vector $D_\beta$ (inhomogeneous vector field contributions are absorbed into $C$). Finally, $\xi^0$ and $\zeta^0$ are the bare piezoaxionic and electroaxionic tensors of Eqs.~\ref{eq:xidef} and~\ref{eq:zetadef}, respectively, while the effective ``axionic charge tensor"
 $W_{\alpha \beta}^s$  is given by:
\begin{align}
W_{\alpha \beta}^s =  4 \pi e \frac{\dd \mathsf{S}}{\dd \overline{\theta}} \frac{\partial}{\partial u_\alpha^{ls}}  \sum_{j,m_j} \left[\epsilon_s \epsilon^*_{p_{j,m_j}} \mathcal{M}_{j,m_j,\beta} + \cc \right]. \nonumber
\end{align}

We can write Eq.~\ref{eq:Ucomp} in a more compact matrix form:
\begin{align}
&U = \label{eq:Umat} \\
&+\frac{1}{2} \vect{S}^\intercal \vect{c}^0 \vect{S} - \vect{D}^\intercal \vect{h}^0 \vect{S} + \frac{1}{2} \vect{D}^\intercal \vect{\beta}^{\infty} \vect{D}  - \hat{\vect{I}}^\intercal \overline{\theta}_a\vect{\xi}^0 \vect{S} - \hat{\vect{I}}^\intercal \overline{\theta}_a \vect{\zeta}^0 \vect{D}\nonumber \\
&+ \frac{1}{N V_c} \left \lbrace \frac{1}{2} \vect{u}^\intercal \vect{C} \vect{u} - e \vect{u}^\intercal \vect{Z} \vect{D} +  \vect{u}^\intercal \vect{G} \vect{S} - e \vect{u}^\intercal \vect{W} \hat{\vect{I}} \overline{\theta}_a \right\rbrace \nonumber, 
\end{align}
with bold type indicating matrix form of the tensors, and index contractions understood through tensor ordering and the transpose ${}^\intercal$. 

We are interested in long-wavelength modes of strain $\vect{S}$ and electric displacement vector $\vect{D}$, with short-wavelength atomic displacements $\vect{u}$ integrated out. These ``short modes'' can be integrated out by requiring that the force on each atom vanishes:
\begin{align}
    0 = \frac{\partial U}{\partial \vect{u}} \quad \Rightarrow \quad \vect{u} = \vect{C}^{-1} \left[e \vect{Z} \vect{D} - \vect{G} \vect{S} + e \vect{W} \hat{\vect{I}} \overline{\theta}_a \right]. \label{eq:usol}
\end{align}
Insertion of this solution back into Eq.~\ref{eq:Umat} yields:
\begin{align}
    &U = \label{eq:Umat2} \\
&+\frac{1}{2} \vect{S}^\intercal \vect{c}^D \vect{S} - \vect{D}^\intercal \vect{h} \vect{S} + \frac{1}{2} \vect{D}^\intercal \vect{\beta}^S \vect{D}  - \hat{\vect{I}}^\intercal \overline{\theta}_a \vect{\xi} \vect{S} - \hat{\vect{I}}^\intercal \overline{\theta}_a \vect{\zeta} \vect{D}, \nonumber
\end{align}
which is identical to Eq.~\ref{eq:int-energy} in the main text, with the effective crystal tensors given by:
\begin{align}
    \vect{c}^D &= \vect{c}^0 - \frac{\vect{G}^\intercal \vect{C}^{-1} \vect{G}}{N V_c}, \\
    \vect{\beta}^S &= \vect{\beta}^\infty - e^2\frac{\vect{Z}^\intercal \vect{C}^{-1} \vect{Z}}{N V_c}, \\
    \vect{h} &= \vect{h}^0 - e \frac{\vect{Z}^\intercal \vect{C}^{-1} \vect{G}}{N V_c}, \\
    \vect{\xi} &= \vect{\xi}^0 - e  \frac{\vect{W}^\intercal \vect{C}^{-1} \vect{G}}{N V_c}, \\
    \vect{\zeta} &= \vect{\zeta}^0 + e^2  \frac{\vect{W}^\intercal \vect{C}^{-1} \vect{Z}}{N V_c}.
\end{align}
$\vect{\xi}$ and $\vect{\zeta}$ are the piezoaxionic and electroaxionic effective tensors after integrating out the high-wavenumber modes. This procedure is analogous to the well-known correspondence for the standard crystal tensors $\vect{c}^D$, $\vect{\beta}^S$, and $\vect{h}$ to their bare counterparts $\vect{c}^0$, $\vect{\beta}^\infty$, and $\vect{h}^0$.

In much of the main text, we follow Voigt notation to describe the constitutive equations and dynamics implied by Eq.~\ref{eq:Umat2}, for clarity, and consistency with literature on piezoelectric crystals. The Voigt prescription reduces the order of symmetric tensors by removing repeated components; for instance, it reduces all $3 \times 3$ symmetric tensors to $6$ dimensional vectors. The strain tensor is given by:
\begin{equation}
    S_{\alpha \beta} = \begin{pmatrix}
S_{xx} & S_{xy} & S_{xz}\\
S_{xy} & S_{yy} & S_{yz} \\ 
S_{xz} & S_{yz} & S_{zz}
\end{pmatrix},
\end{equation}
and is therefore simplified to the 6-dimensional vector:
\begin{align}
S_i& = \left(S_{xx}, S_{yy}, S_{zz}, S_{yz}, S_{xz}, S_{xy} \right)  \\
   & = \left(S_1, S_2, S_3, S_4, S_5, S_6 \right),
\end{align}
such that the scalar product is preserved, i.e.:
\begin{eqnarray}
\sum_{\alpha, \beta} S_{\alpha \beta} S_{\alpha \beta} = \sum_i S_i S_i.
\end{eqnarray}
Vectors (1-tensors) such as $D_\beta$ still have 3 components, while 4-tensors and 3-tensors reduce to $6\times6$ and $6\times 3$ ``2-tensors'' (in the Voigt convention), respectively. It is now straightforward to rewrite the constitutive Eqs.~(\ref{eq:Eelecfield}), and~(\ref{eq:Tstress}) in Voigt notation as Eqs.~(\ref{eq:EVoigt}) and (\ref{eq:TVoigt}) in Sec.~\ref{sec:crystal}, for example.

\section{Piezoelectric Equivalent Circuit Components}\label{sec:piezoequiv}
To separate the immittance of the piezoelectric crystal into constituent electrical components, we make use of the power series \cite{ballato2001}:
\begin{equation}
    \tan{x} = \sum^{\infty}_{j=1}\frac{8x}{(2j-1)^2\pi^2-4x^2}.
\label{eq:tan_series}
\end{equation}
Each subsequent term in the expansion corresponds to an overtone of the mechanical resonance. In the vicinity of the fundamental resonance, we can approximate the immittance by the first term in the series. For modes whose electric field is along the direction of acoustic wave propagation, e.g. Eq.~\ref{eq:Zcrystal} this gives an expression for the impedance of the form:
\begin{equation}
    Z = \frac{1}{i \omega C_c}\left( 1-k^2 \frac{8}{\pi^2-4(\omega \ell/ 2 v)^2} \right).
\label{eq:impedance_thickness_modes}
\end{equation}
Equation~\ref{eq:impedance_thickness_modes} is equivalent to the following electrical components, arranged as in Fig.~\ref{fig:equivalentcircuit}
\begin{align}
    C_m &= \frac{8C_c k^2}{\pi^2-8k^2}, \label{eq:Cm} \\
    L_m &= \frac{\ell^2}{8C_c \,k^2v^{ 2}}. \label{eq:Lm}
\end{align}
To take into account mechanical losses, we add a resistor to the circuit whose value is set by the mechanical quality factor $Q_m$ of the crystal: 
\begin{equation}
    R_m=\frac{\pi\, \ell}{8 \,C_c\, k^2\,v}\frac{1}{Q_m}. \label{eq:Rm}
\end{equation}
We reiterate that we do not use the circuit elements of Eqs.~\ref{eq:Cm}--\ref{eq:Rm} in the analysis in the main text---we use the exact expression of Eq.~\ref{eq:Zcrystal} with imaginary crystal tensors. They are shown here to illustrate that near any mechanical resonance frequency, the behavior of the crystal can be accurately described by ``standard'' circuit elements. 

In App.~\ref{app:piezomodes}, we include additional modes that could be used in our setup. For modes such as the whose electric field is in the perpendicular direction to the acoustic excitation, using Eq.~\ref{eq:tan_series} gives an admittance of the form:
\begin{align}
    Y &= \frac{1}{Z} = i \omega C_c\left( 1+K^2 \frac{8}{\pi^2-4(\omega \ell/ 2 v)^2} \right).
\label{eq:admittance_lateral_modes}
\end{align}
The equivalent circuit components are given by:
\begin{align}
    C_m &= \frac{8C_c K^2}{\pi^2}, \label{eq:lateralfieldcomps1} \\
    L_m &= \frac{\ell^2}{8C_c \,K^2v^{ 2}},  \label{eq:lateralfieldcomps2}\\
    R_m &=\frac{\pi\, \ell}{8 \,C_c\, K^2\,v}\frac{1}{Q_m}.  \label{eq:lateralfieldcomps3}
\end{align}

\section{More Modes}\label{app:piezomodes}
In the main text (Sec.~\ref{sec:experiment}), we introduced the thickness expander mode, which is well understood and known to possess high mechanical quality factors in cryogenically cooled quartz and similar materials~\cite{Galliou2015, goryachev:tel-00651960}. Here we present two more modes that could be relevant for our setup. These have been constructed with the symmetry group of quartz, $32$, in mind, but can be easily adapted to crystals of other symmetry groups.  Much of the analysis in this appendix is based on the modes presented in \cite{berlincourt1964piezoelectric}.

First, we give an example of a length expander mode, which could be useful at low frequencies. Since the mechanical resonance frequency of this mode is set by the (longer) length dimension of a crystal bar rather than the thickness dimension of a thin plate, it could probe lower frequencies using a crystal of similar size compared to a thickness mode. The length expander mode, however, typically has a lower mechanical Q-factor than the thickness mode~\cite{zelenka1986}.

The second mode we include in this appendix is a thickness mode of a thin plate with a lateral electric field. Like the thickness mode in the main text, the acoustic excitations propagate in the thickness direction of the crystal, i.e. the shortest dimension of the thin plate, but the electrodes are now placed on the minor faces so that the direction of electrical excitation is perpendicular to the thickness direction instead of parallel. This mode is less well explored experimentally but has some theoretically promising features: the axion induced voltage is integrated over the length of the plate rather than the thickness direction, which could increase the signal size by around an order of magnitude, while in principle having similar mechanical losses to the standard thickness mode. In addition, the electromechanical resonance frequency of the crystal coincides with the resonant frequency of the signal, unlike in the parallel field case where the signal frequency coincides with antiresonance, which could also improve the sensitivity of the setup and make frequency scanning simpler. 

Another useful measure for comparing the potential effectiveness of different modes is the electromechanical coupling factor $k$, which expresses the proportion of electrical energy that can be converted into mechanical energy by the crystal, or vice versa, and can be seen as a measure of the strength of electro-elastic interactions of a given mode \cite{berlincourt1964piezoelectric}.  High coupling factors suggest a larger piezoaxionic signal due to efficient conversion of mechanical energy, and can also be seen directly from Eq.~\ref{eq:Rm} to lower the mechanical losses of the system. We will see that the new modes here have coupling factors that are comparable to the original length expander mode, which has $k
\sim 0.09$ for quartz with parameters as given in \cite{yang2006analysis}.

 We focus on longitudinal modes rather than shear modes since these develop larger quality factors at very low temperatures \cite{goryachev2011, Goryachev2012}. Nevertheless, we briefly mention here that one of the most commonly used modes in quartz is the thickness shear mode, which has the largest coupling factor $k\sim0.14$. To obtain analogous expressions for $V_a$, $Z_\mathrm{crys}$, and $Q$ for the thickness shear mode, a similar analysis to Sec.~\ref{sec:experiment} applies if ones takes the thickness,  wavenumber, and electric field to be aligned in the 2-direction ($\ell_2 \ll \ell_1, \ell_3$, $\vect{k} \propto \hat{\vect{x}}_2$, and $\vect{E} \propto \hat{\vect{x}}_2$), and the displacement in the 1-direction ($\vect{u} \propto \hat{\vect{x}}_1$). The constitutive equations then take the same form as Eqs.~\ref{eq:T1constitutive},\ref{eq:E1constitutive} with the replacements: $T_1 \to T_6$, $S_1 \to S_6$, $E_1 \to E_2$, $D_1 \to D_2$,  $\hat{I}_1 \to \hat{I}_2$, $c_{11}^D \to c_{66}^D$, $h_{11} \to h_{26}$,  $\beta_{11}^S \to \beta_{22}^S$, $\xi_{11} \to \xi_{62}$,  and $\zeta_{11} \to \zeta_{22}$.

\subsubsection*{Length expander mode in bar with the axion-induced electric field parallel to length} 
Consider a narrow bar with its length along the $x_1$-direction, and cross-sectional dimensions that are small compared to its length:  $\ell_2,\ell_3<\ell_1$. The electrodes are placed on the faces normal to the $x_1$-direction. Like with the thickness modes, $D_2 = D_3 = 0$ and $D_1$ is spatially uniform. Negligible cross-sectional dimensions gives only $T_1 \neq 0$. We therefore choose $D$ and $T$ as independent variables, which suggests using the following alternative form of the constitutive equations \cite{berlincourt1964piezoelectric}:
\begin{align}
S_1 &= +s_{11}^D T_1 + g_{11} D_1 + \xi'_{11}\hat{I}_1\overline{\theta}_a, \\ 
E_1 &= -g_{11}T_1 + \beta^T_{11} D_1 - \zeta^{\,T}_{11} \hat{I}_1\overline{\theta}_a.
\end{align}
We see that the nuclear spins should be polarized in the $x_1$ direction. The matrices of proportionality constants, written in terms of those contained in the internal energy of Eq.~\ref{eq:Umat}, are given by:
\begin{align}
    \vect{s}^D &= (\vect{c}^D)^{-1}, \\
    \vect{g} &= \vect{h}\, \vect{s}^D, \\
    \vect{\beta}^T &= \vect{\beta}^S - \vect{h}\, \vect{s}^D\, \vect{h}^\intercal, \\
    \vect{\xi}' &= \vect{s}^D\, \vect{\xi}, \\ 
    \vect{\zeta}^T &= \vect{\zeta} + \vect{h} \,\vect{s}^D\, \vect{\xi};
\end{align}
where superscripts $^T$ and $^D$ denote tensors defined at constant stress or electric displacement (and superscript $^\intercal$ once again denotes a transpose). 

The equation of motion for this  mode is $\rho \ddot u_1 = \partial_1 T_{1} =\partial_1^2 u_1 / s_{11}^D$. The solution to this equation that also satisfies the boundary conditions at the free surfaces $T_1 = 0$ at $x_1 = (0, \ell_1)$ is:
\begin{align}
u_1 &= \frac{g_{11} D_1 + \xi_{11}^{'} \hat{I}_1 \overline{\theta}_a}{\omega/v^D}  \\
&\phantom{=} \times \left[\sin  \frac{\omega x_1}{v^D}  - \tan\frac{\omega \ell_1}{2v^D}  \cos\frac{\omega x_1}{v^D}  \right], \nonumber
\end{align} 
with $v_b^D = \sqrt{\frac{1}{\rho s_{11}^D}}$ the crystal sound speed for a bar with constant-D conditions.
Above, the quantities $u_1$, $D_1$, and $\hat{I}_1$ are again assumed to be oscillatory with angular frequency $\omega$.

Rearranging the constitutive equation for the component $E_1$ in terms of $S_1$ gives:
\begin{equation}
E_1 = -\frac{g_{11}}{s_{11}^D}S_1+\left( \frac{g_{11}^2}{s_{11}^D}+\beta_{11}^T\right) D_1 + \left( \frac{g_{11} \xi_{11}^{'}}{s_{11}^D}+\zeta_{11}^{T}\right) \hat{I}_1 \overline{\theta}_a,
\end{equation}
which can be integrated over the length of the crystal to find the voltage across the electrodes like for the thickness modes:
\begin{align}
V &= \int_0^{\ell_1} \dd x_1 \, E_1 = Z I + V_a, \\
V_a &= -\left[ \frac{g_{11} \xi^{'}_{11}}{s_{11}^D} \frac{2 v^D}{\omega}  \tan \frac{\omega \ell_1}{2 v^D} - \ell_1 \left(\frac{g_{11} \xi^{'}_{11}}{s_{11}^D} +   \zeta_{11}^{T} \right)   \right] \hat{I}_1 \overline{\theta}_a, \\
Z &= \frac{1}{i\omega C_c} \left[ 1 - k^2 \frac{2 v^D}{\omega \ell_1} \tan \frac{\omega \ell_1}{2 v^D}\right], \\
C_c &= \frac{\ell_2 \ell_3}{\ell_1}\left( \frac{g_{11}^2}{s_{11}^D} + \beta_{11}^{T}\right)^{-1},\\
k &= \frac{g_{11}}{\sqrt{g_{11}^2+\beta_{11}^T s_{11}^D}}.
\label{eq:signalparameters1}
\end{align}
For quartz, the coupling factor $k$ of this mode is $\sim 0.10$, which is comparable to the thickness expander mode in the main text.

\subsubsection*{Thickness expander mode with axion-induced electric field perpendicular to thickness}
Like in Sec.~\ref{sec:experiment}, we will again take the piezoelectric crystal to be a rectangular prism with side lengths $\ell_i$ of high aspect ratio (thin plate): $\ell_2 \ll \ell_1, \ell_3$. Notice this time we have oriented the crystal slightly differently, with $l_2$ rather than $l_1$ being the thickness dimension. The electroded faces are normal to the $x_1$-direction, which leads to the boundary condition that on the electroded faces $E_2=E_3=0$. The electroded surfaces are also equipotential surfaces, meaning that $E_1$ is independent of $x_2$, i.e. $\partial E_1/\partial x_2=0$ The plate is considered to be laterally clamped, so that only $S_2 \neq 0$, with the other 5 strains vanishing identically. On the free surfaces at $x_2=(0,l_2)$, the stress $T_2=0$. These conditions suggest that the independent variables are chosen to be $S$ and $E$, which leads to the constitutive equations:
\begin{align}
    T_2 &= +c^E_{22}S_2 - e_{12} E_1 -\xi^E_{21} \hat{I}_1 \overline{\theta}_a, \\ 
    D_1 &= +e_{12} S_2 + \epsilon^S_{11} E_1 - \zeta'_{11} \hat{I}_1 \overline{\theta}_a.
\end{align}
The nuclear spins are polarized in the $x_1$ direction. The Voigt matrices of proportionality constants are given by:
\begin{align}
    \vect{c}^E &=  \vect{c}^D - \vect{h}^\intercal \vect{\epsilon}^S \vect{h},\\
    \vect{\epsilon}^S &= (\vect{\beta}^S)^{-1}, \\
    \vect{e} &= \vect{\epsilon}^S \vect{h},\\
    \vect{\xi}^E &= \vect{\xi} + \vect{h}^\intercal \vect{\epsilon}^S \vect{\zeta},\\
    \vect{\zeta'} &= \vect{\epsilon}^S \vect{\zeta}.
\end{align}
The wave equation for this  mode is $\rho \ddot u_2 = \partial_2 T_{2} = c_{22} \, \partial_2^2 u_2$. As usual, we take harmonic factors to be implicit. 
The solution to the wave equation that satisfies the boundary conditions listed above is given by
\begin{align}
u_2 &= \frac{e_{12}E_1 + \xi^E_{21}  \hat{I}_1 \overline{\theta}_a}{c^E_{22} \omega/v^E}  \\
&\phantom{=} \times \left[\sin  \frac{\omega x_2}{v^E}  - \tan\frac{\omega \ell_2}{2v^E}  \cos\frac{\omega x_2}{v^E}  \right], \nonumber
\end{align} 
with $v^E = \sqrt{\frac{c_{22}^E}{\rho}}$ the crystal sound speed for a bar with constant-E conditions.

We can now find the current in the piezoelectric by integrating the time derivative of the electric displacement over its thickness:
\begin{equation}
I = \ell_3 \int_0^{\ell_2} \dot{D_1}\, \dd x_2,
\end{equation}
while the voltage across the electrodes is given by the integral over its length:
\begin{equation}
V = \int_0^{\ell_1} E_1 \,\dd x_1,
\end{equation}
this gives the results:
\begin{align}
I &= \frac{V}{Z} + I_a, \\
I_a &= i \omega \ell_3 \left[ \frac{e_{12}\, \xi^E_{21}}{c^E_{22}}\frac{2 v^E}{\omega} \tan{\frac{\omega\, l_2}{2 v^E}} - \zeta'_{11} l_2 \right] \hat{I}_1 \overline{\theta}_a, \\
\frac{1}{Z} &= i \omega C_c \left[ 1 + k^2 \frac{2 v^E}{\omega \ell_2} \tan \frac{\omega \ell_2}{2 v^E}\right] ,\\
C_c &= \frac{\ell_2 \ell_3}{\ell_1}\varepsilon_{11}^S, \\
k &= \frac{e_{12}}{\sqrt{c^E_{22}\epsilon^S_{11}}};
\end{align}
where we see that the axion signal enters as a current source in parallel with the crystal.
The coupling factor $k$ in quartz is almost identical to the other thickness expander mode, with both having $k\sim 0.09$ for quartz parameters.
Converting to an equivalent voltage source gives:
\begin{align}
V_a = I_a Z &= \frac{\ell_1}{\epsilon_{11}^S} \left\lbrace 1+ k^2 \frac{2 v^E}{\omega \ell_2} \tan \frac{\omega\, \ell_2}{2 v^E} \right\rbrace^{-1} \\
&\phantom{=} \times \left[   \frac{e_{12}\, \xi^E_{21}}{c_{22}^E}  \frac{2 v^E}{\omega \ell_2} \tan 
\frac{\omega \,\ell_2}{2 v^E} -\zeta'_{11} \right] \hat{I}_1 \overline{\theta}_a. \nonumber
\end{align}
At the mechanical resonance frequency of the plate, $f = \frac{2 v^E}{l_2}$, the admittance $1/Z$ diverges, implying that the plate is also at electromechanical resonance (assuming it is not yet loaded by external electrical components). The equivalent circuit components can be found using Eqs.~\ref{eq:lateralfieldcomps1}, \ref{eq:lateralfieldcomps2} and \ref{eq:lateralfieldcomps3}. 

The additional modes presented in this appendix suggest that by changing the orientation of the crystal, the direction of nuclear spin polarization and the placement of electrodes, one could improve the frequency range and sensitivity of the experiment when subject to limitations in number of crystals and their size. A full optimization analysis including a larger variety of crystal modes such as these will be left to future work.

\end{document}